\documentclass[useAMS,usenatbib,twocolumn]{mnras}

\newcommand{\tpara}{\delta\theta_{\parallel}}
\newcommand{\tperp}{\delta\theta_{\perp}}

\newcommand{\tparaL}{\langle\tpara\rangle}
\newcommand{\tperpL}{\langle\tperp\rangle}


\usepackage[T1]{fontenc}
\usepackage{aecompl}
\usepackage{hyperref}
\usepackage{graphicx} 
\usepackage{times} 
\def\aj{AJ}
\def\apj{ApJ}
\def\apjl{ApJL}
\def\mnras{MNRAS}
\def\nat{\rm Nature}
\def\prd{PhysRevD}
\def\araa{ARA\&A}
\def\aap{A\&A}
\def\apss{Ap\&SS}

\begin{document}

\title[The Shadow Bar]
 {Dark Matter Trapping by Stellar Bars: The Shadow Bar}

 \author[M. Petersen et al.]
 {Michael~S.~Petersen,$^1$\thanks{mpete0@astro.umass.edu}
   Martin~D.~Weinberg,$^1$ Neal~Katz$^1$ \\ $^1$University of
   Massachusetts at Amherst, 710 N. Pleasant St., Amherst, MA 01003}

\maketitle

\begin{abstract}
  We investigate the complex interactions between the stellar disc and the dark-matter halo during bar formation and
  evolution using N-body simulations with fine temporal resolution and
  optimally chosen spatial resolution.  We find that the forming
  stellar bar traps dark matter in the vicinity of the stellar bar
  into bar-supporting orbits. We call this feature the \emph{shadow
    bar}. The shadow bar modifies both the location and magnitude of
  the angular momentum transfer between the disc and dark matter halo and
  adds 10 per cent to the mass of the stellar bar over 4 Gyr.  The
  shadow bar is potentially observable by its density and velocity
  signature in spheroid stars and by direct dark matter detection experiments.
  Numerical tests demonstrate that the
  shadow bar can diminish the rate of angular momentum transport from
  the bar to the dark matter halo by more than a factor of three over
  the rate predicted by dynamical friction with an untrapped dark halo, and thus provides a
  possible physical explanation for the observed prevalence of fast
  bars in nature.
\end{abstract} 


\section{Introduction} \label{sec:introduction}

Galactic bars change the internal structure of disc galaxies by
rearranging the angular momenta and energy of orbits that would 
otherwise be conserved.  As many as two-thirds of local galaxies show bar
features in the infrared \citep{sheth08}. Bars have been suggested to
affect both structural properties such as bulge structure
\citep{kormendy04,laurikainen14} and stellar breaks
\citep{munoz13,kim14}, as well as evolutionary properties such as star
formation rates \citep{masters12,cheung13} and metallicity gradients
\citep{williams12}.  \citet{weinberg02} also showed that bars can affect the
central profiles of dark matter haloes.
Bars are the strongest of the persistent disc
asymmetries and, therefore, provide a dynamical laboratory for
understanding the long-term evolution of galaxies.

For simplicity of analysis, many previous theoretical works have adopted a
component-by-component analysis of angular momentum transport by
adopting a fixed halo potential, a rigid bar, or both.  However, the
baryonic disc and dark matter halo comprise a single system that
shares the same dynamics as a consequence of their mutually generated
gravitational field.  Therefore, some fraction of the dark-matter
orbital elements will overlap with those in the disc and, thus,
there is little reason not to suspect that some of the dark matter halo
will respond similarly to the stellar disc.  The dynamical
interplay between these components is of considerable interest in
developing a comprehensive understanding of the disc-halo system.\\

In this paper, we present a time-dependent analysis of trapped stellar
and halo orbits in a fully self-consistent simulation.  We find
evidence for the formation and subsequent secular evolution of a
trapped population of halo orbits, a \emph{shadow bar}, that arises in
response to the same collective dynamics that make the classic stellar
bar. The shadow bar fundamentally changes the orbital structure of the
halo in the vicinity of the bar.  These same orbits, when unperturbed
by the bar, would contribute strongly to the disc-halo torque responsible
for slowing the bar, which can be understood in the context of simple
perturbation theory models. We investigate and document the magnitude
of halo trapping, finding that the trapping affects both the
dark-matter density and velocity structure.

A well-studied consequence of angular momentum transport in barred
galaxies is the tension between the theorised slowing of the bar
pattern speed \citep{tremaine84,weinberg85} and observations that do
not show a slowing pattern speed (e.g. \citealt{aguerri15}). This is
often interpreted as evidence for only small amounts of dark matter at the
centre of galaxies \citep{debattista00,sellwood06b,villavargas09} or
as a motivation for changing the law of gravity
\citep{tiret07}. However, the dark-matter orbits trapped by the bar
are unavailable for secular evolution, significantly reducing the halo
torque on the bar.

For our own Milky Way (MW)---a barred galaxy---several experiments may
detect the influence of a shadow-bar modified dark matter density and
velocity distribution. These include ongoing analysis of stellar
surveys such as RAVE \citep{steinmetz06}, GAIA \citep{gilmore12}, and
APOGEE \citep{allende08} since halo stars might have similar orbits to dark
matter particles. Some of these studies have already seen suggestions
of rotation in a bulge-bar component (\citealt{rojas14},
\citealt{soto14}).  Signs of a shadow bar could also affect the 
interpretation of tentative signals from direct dark matter detection experiments,
e.g. CoGeNT \citep{aalseth13}, CDMS II \citep{agnese13}, and DAMA/LIBRA
\citep{bernabei14}.

\begin{figure}
  \centering 
  \includegraphics[width=3.5in]{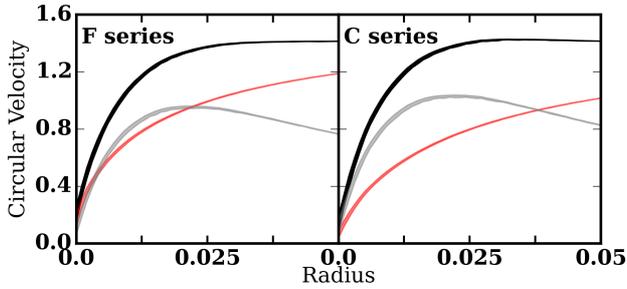} 
  \caption{Initial circular velocity curves for the F-series cusp models (left
    panel) and C-series cored models (right panel). The circular velocity from
    $p=0.2$ to $p=0.8$ is shown in black where $p$ is the quantile
    value for the rank-ordered circular velocities. The $p=0.2$ to $p=0.8$ contributions from the
    disc (grey line) and halo (red line) have been separated to
    demonstrate the maximality of the disc. Radius and
    circular velocity are in virial units (see
    Section~\ref{subsec:methods}), where a radius of $0.01$
    coresponds to $3$
    kpc and a circular velocity of $1.0$ corresponds to 150~km s${}^{-1}$ when scaled to
    a Milky Way mass galaxy. \label{fig:rotcurve}}
\end{figure}

This paper is organised as follows: the next section describes the
initial conditions for our fiducial simulation (section
\ref{subsubsec:fiducial}), a series of simulations with varying halo
properties (section \ref{subsubsec:halovariants}), and the details of
our N-body solver (section \ref{subsec:integration}).  We include two
idealised numerical experiments with modifications to the bar
perturbation to better understand the importance of the bar growth
rate on halo orbit trapping and halo structure.  Our results are presented in section
\ref{sec:results}.  We examine the fiducial simulation in detail,
beginning with an overall summary picture of the fiducial simulation
in section \ref{subsec:orbitstat}, followed by studies of the disc and
halo orbit morphologies in sections \ref{subsubsec:discorbits} and
\ref{subsubsec:haloorbit}, respectively. These analyses reveal that in
addition to the stellar bar, a significant mass fraction of the halo
within a bar length is trapped into the bar potential. The dynamical
properties of the entire trapped orbit population are discussed in
section \ref{subsec:dynquan} and contextualised using perturbation
theory.  This is followed by an exploration of the overall dynamical
consequences of the shadow bar itself in section
\ref{subsec:effect}. Section \ref{sec:discussion} connects with
previous work by using a suite of experiments motivated by published
models and further demonstrates the robust nature of the shadow bar.
We conclude with a summary and discussion of the stellar disc--dark
matter halo relationship and our detailed orbital analysis in section
\ref{sec:conclusion}.

\section{Methods}  \label{sec:fiducial} 

\subsection{Initial conditions} \label{subsec:methods} 

This section motivates our mass models and describes the realisation
of the initial conditions.  We begin with the details of our fiducial
simulation.  The phase space is realised using a methodology that
closely follows \citet{holleyb05}, hereafter HB05. We then discuss
model variants that (1) address typical literature treatments; (2)
seek to model physical processes in the universe; and (3) verify the
theorised dynamical implications of the fiducial experiment. Our key
simulations are summarised in Table \ref{tab:simlist}.

In all the simulations presented here, the number of disc particles
and halo particles are $N_{\rm disc}=10^6$ and $N_{\rm halo}=10^7$,
respectively. The disc particles have equal mass.  The halo particles
have a number density $n_{\rm halo}\propto r^{-\alpha}$, where
$\alpha=2.5$.  This steep power law better resolves the inner halo in
the vicinity of the disc bar, our subject of interest.  The particle's
masses are assigned to simultaneously reproduce the desired halo mass
density and the number density. In practice, the halo resolution is
improved by a factor of 100 within a disc scale length
($M_{r<R_d}=0.005M_{\rm vir}$, $N_{r<R_d}=5\times10^5$) compared to
using a fixed halo particle mass. The effective halo particle
number within all disc-relevant radii is then $N_{\rm halo}\sim 10^9$,
more than satisfying the criteria of \citet{weinberg07b} for an
adequate treatment of resonant dynamics.

All units are scaled to so-called \emph{virial} units with $G=1$ that
describe the mass and radius of the halo at the point of halo collapse
in the cold-dark-matter scenario. For comparison to the MW, we follow
the estimated halo mass of \citet{boylan13} such that $M_{\rm
  vir}=1.6\times10^{12} ~{\rm M}_\odot$, making $R_{\rm vir}=1.0$
correspond to 300 kpc, a velocity of 1.0 equivalent to 150 km
s$^{-1}$, and a time of 1.0 equivalent to 2.0 Gyr. The system is
constructed such that $M(R_{\rm vir})=1.0$.

\subsubsection{Fiducial simulation}\label{subsubsec:fiducial}

We follow the now standard prescription for producing a strong bar in
an N-body simulation by choosing a fiducial model with a slowly rising
rotation curve. This
simulation is labelled F in Table \ref{tab:simlist}, using the initial
rotation curve shown by the F-series rotation curve in Fig.~\ref{fig:rotcurve}
(left panel). The contributions from the individual components are
derived by calculating the radial force contributions from each
component separately: $v_{c,{\rm disc}}^2=rF_{r,{\rm disc}}$
and $v_{c,{\rm halo}}^2=rF_{r,{\rm halo}}$ where $r$ is the radius of
the particle, and $F_r$ is the radial force contribution from the
corresponding component at
the particle's position. The slowly rising rotation curve minimises the
bar-damping effect of the initial inner Lindblad resonance and allows
the region inside of corotation to act as a ``resonant cavity'' for
bar formation \citep{sellwood93}. For a rising rotation curve, one can
show using first-order perturbation theory that the response of a
nearly circular orbit to the bar perturbation is to slow and even
reverse its apse precession rate (the so-called ``donkey effect'' of
\citealt{lynden72}, see \citealt{binney08}).  This results in the orbit
lingering near the position angle of the bar, causing the effective
bar strength, the quadrupole perturbation strength specifically, to
increase.  This growth extends the reach of the perturbation, inducing
more orbits to slow or reverse their precession rate, resulting in
an exponential growth for the bar. The fiducial simulation is analysed in
section \ref{sec:results}.

The fiducial simulation has an initially exponentially-distributed
stellar disc embedded in a fully self-consistent non-rotating $c =
R_{\rm vir}/r_s \approx15$ NFW dark-matter halo \citep{navarro97}
where $r_s$ is the scale radius, and $R_{\rm vir}$ is the virial
radius:
\begin{equation} \label{eq:haloprof}
\rho(r) = \frac{\rho_0r_s^3}{(r+r_c)(r+r_s)^2}
\label{eq:nfw}
\end{equation}
where $\rho_0$ is set by the chosen mass and $r_c=0$ is the core radius for the fiducial
model.  The dynamical implications of a core produced by early
evolutionary processes (\citealt{weinberg02}, \citealt{sellwood09})
will be explored in section \ref{sec:results} and \ref{sec:discussion}
by comparison to simulations with non-zero values of $r_c$.  The
halo velocities are realised from the distribution function produced
by an Eddington inversion of $\rho(r)$ (see \citealt{binney08}).  Eddington inversion provides
an isotropic distribution, roughly consistent with the observed
distributions in dark-matter only $\Lambda$CDM simulations. This
method naturally results in a non-rotating halo; we will describe
realising rotating haloes in section \ref{subsubsec:halovariants}.

For all the simulations here, we adopt an exponential radial stellar disc profile
with an isothermal vertical distribution:
\begin{equation}\label{eq:discprof}
\rho_d(R,z) = \frac{M_{\rm d}}{8\pi z_0R_d^2} e^{-R/R_d}
{\rm sech}^2 (z/z_0)
\end{equation}
where $M_d$ is the disc mass, $R_d$ is the disc scale length, and
$z_0$ is the disc scale height. We set $M_d=0.025$, $R_d=0.01$, and
$z_0=0.001$ throughout. All simulations in this work have a disc:halo mass ratio of 1:40 inside of
its virial radius\footnote{To preserve equilibrium, the halo extends
  beyond a virial radius, where it is truncated, but here we consider
  the mass within a virial radius for this ratio calculation.}; while
a comparatively low mass ratio for studies in the literature (e.g. contrasting with
\citealt{debattista00}; \citealt{athanassoula02}; \citealt{saha12}),
we choose this mass ratio to
mimic the mass ratio for the $z=0$ MW, as determined through
simulations \citep{vogelsberger14}, abundance matching
(\citealt{kravtsov13}; \citealt{moster13}), and stellar kinematics
\citep{bovy13}. The disc parameter values are consistent with the MW
values of \citet{bovy13}, who found $M_\star\approx4.6\times10^{10}$,
or $0.029~M_{\rm vir}$ (again using the scaling from
\citealt{boylan13}); $R_d=2.15$ kpc and $z_0=370$ pc, which are
reasonably similar to our values. We do not include a bulge due to the
uncertainty in its phase-space distribution, but this should not limit
the application of our models to the observed universe since the
dynamical mechanisms governing bar formation and evolution are
dominated by the graviatational potential outside the bar region.
Similarly for the thick disc, although we will address the dynamical
effects of a thick disc in a later paper. We additionally do not include a
gas component. In the present-day, gas is a tracer component in the MW;
while the gas fraction may have been higher in the past, this quantity
is highly uncertain. Literature results have demonstrated scenarios
where gas accretion can destroy and re-form the bar over a
cosmological \citep{bournaud02,bournaud05}. However, other
simulations including gas as part of the initial conditions
\citep{villavargas10,athanassoula13} do not show signs
of bar destruction, though the bars resulting from gas-rich
simulations may be weaker. While including gas as part of the initial
conditions may slow the
rate at which bars form, the presence of a gas component does not fundamentally alter the
dynamics. In light of this, we suggest that our models are adequate
for learning about the dynamical mechanism of DM trapping, and believe
that the results presented here would remain largely unchanged in the
presence of a collisional component with present-day parameters.

The F-series rotation curve has a disc fraction $f_D\equiv
V_{c,\star}/V_{c,{\rm tot}}=0.71$ at $R=2.2R_d$, the radius at which
the exponential disc reaches the maximum circular
velocity\footnote{The parameter $f_D$ is often called the
  `maximality'. By the nature of our simulations, $f_D$ only measures
  the contribution of the stellar component rather than the total
  baryonic component, so we quote literature results which are able to
  isolate stellar contributions from the full baryonic contributions.}. \citet{martinsson13} found that for typical spiral galaxies,
$f_D=0.4-0.7$, with $\langle f_D\rangle=0.57$, making our disc more
maximal than the typical spiral. \citet{piffl14a} measured $f_D=0.63$
for the MW, in agreement with the findings of
\citet{martinsson13}. However, \citet{bovy13} suggested that the MW
may be nearer to a maximal disc, with $f_D=0.83$. All methods to
observationally determine $f_D$ rely on assumptions about the
parameters to describe the disc, as well as the structure of the dark
matter halo, hence we believe that within the uncertainties, our fiducial
model
represents a realistic galaxy. The purpose of the
simulation was not to create an exact MW analogue, but to make a cosmologically realistic galaxy.

Disc velocities are chosen by solving the
Jeans' equations in cylindrical coordinates in the combined disc--halo
potential. We set the radial velocity dispersion from the Toomre
stability equation,
\begin{equation}
\sigma_r(R) = Q\frac{3.36G\Sigma(R)}{\kappa(R)},
\end{equation}
where $G=1$, $\Sigma(R)$ is the stellar surface density, and $\kappa$,
the epicyclic frequency (also sometimes written as $\Omega_r$), is given by
\begin{equation}
\kappa^2(R) = R\frac{d\Omega (R)^2}{dR}+4\Omega(R)^2
\end{equation}
where $\Omega$ is the azimuthal frequency (sometimes explicitly
written as $\Omega_\phi$ in cylindrical coordinates). We choose $Q=0.9$ to
facilitate comparisons to the literature
(e.g. \citealt{athanassoula02}) and to ensure the relatively rapid growth
of a bar\footnote{In the absence of a dark matter halo, $Q=0.9$ is
  formally unstable; in the presence of the dark matter halo, this
  value results in a locally stable disc that forms a bar in several
  dynamical times.}.  This departs from our previous choice in HB05 of
a \emph{warm} disc with $Q=1.4$.  Also, we replace the
axisymmetric velocity ellipsoid generated from the epicyclic
approximation in the disc plane used in HB05 by the Schwarzschild
velocity ellipsoid \citep{binney08}. In practice, this latter velocity
ellipsoid improves the initial equilibrium by using the second moment of the
cylindrical collisionless Boltzmann equation with an asymmetric drift
correction:
\begin{equation}
  \sigma_\phi^2 = \left(\frac{\sigma_r\kappa(R)}{2\Omega(R)}\right)^2.
\end{equation}
As in HB05, the vertical velocity dispersion $\sigma_z$ is
\begin{equation}
  \sigma_z^2(R) = \frac{1}{\rho_d(R,z)}\int_z^\infty
  \rho_d(R,z)\frac{\partial\Phi_{\rm tot}}{\partial z}dz.
\end{equation}

From a dynamical standpoint, our relatively low disc-to-halo mass
ratio accomplishes two additional goals: (1) it decreases the strength
of local instabilities that allows us to focus on secular
dynamics; and (2) it enables the exploration of the \emph{slow mode} of
bar growth \citep{polyachenko96}, seemingly not achievable given
\emph{maximal} discs. We examine the former in this paper and will
explore the latter in more detail in a forthcoming paper.

\begin{figure}
  \centering 
  \includegraphics[width=3.5in]{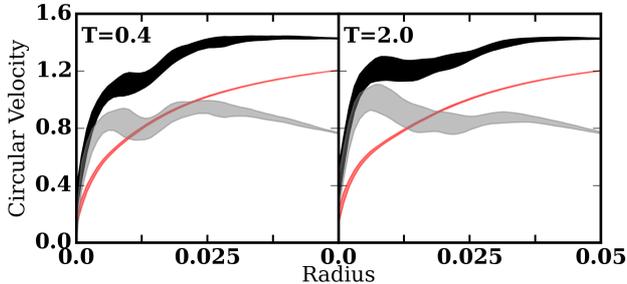} 
  \caption{Circular velocity curves for the fiducial model at system time
    T=0.4 (left
    panel) and T=2.0 (right panel). The circular velocity from
    $p=0.2$ to $p=0.8$ is shown in black. The $p=0.2$ to $p=0.8$ contributions from the
    disc (grey line) and halo (red line), where $p$ is the quantile
    value for the rank-ordered circular velocities, have been calculated separately to
    demonstrate the evolution of each component. Radius and
    circular velocity are in virial units, as in
    Figure~\ref{fig:rotcurve}. System time is also given in virial units,
    where T=1.0 corresponds to 2.0 Gyr when scaled to the MW.
    \label{fig:rotcurve_evolved}}
\end{figure}

\begin{table*} \centering

\begin{tabular}{cccccc} Simulation & scale length ($R_d$) & scale
                                                            height
                                                            ($h$) &
                                                                    disc mass & halo profile & Notes\\ 
\hline \hline 
F & 0.01 & 0.001 & 0.025 & cusp &\\ 
Fr & 0.01 & 0.001 & 0.025 & cusp, $\lambda=0.03$ &\\ 
C & 0.01 & 0.001 & 0.025 & cored, $r_c=0.02$ & \\ 
Cr & 0.01 & 0.001 & 0.025 & cored, $r_c=0.02$, $\lambda=0.03$ &\\ 
Ff & 0.01 & 0.001 & 0.025 & cusp & fixed disc potential\\
Fs & 0.01 & 0.001 & 0.025 & cusp &shuffled halo\\ 
\end{tabular} \caption{List of simulations\label{tab:simlist}.} \end{table*}

\subsubsection{Halo model variants} \label{subsubsec:halovariants}

We investigate two additional classes of halo models based on our
fiducial one: (1) halo models with a cored central profile instead of
a cusp\footnote{A cored central profile satisfies $\rho(r) \propto
  r^{-\alpha}$ where $\alpha\to0$ as $r\to0$; the unmodified NFW has a cuspy
  central profile with $\alpha=1$.}; and (2) rotating haloes. These
variants are inspired both by cosmological hydrodynamic simulations
\citep{bullock01}, and a desire to connect with previous literature
choices.

We construct cored haloes by setting $r_c=0.02$ in equation
(\ref{eq:haloprof}). The halo mass is matched between the cusp and core
model haloes at the virial radius to maintain the disc:halo mass ratio
within the virial radius. This model uses the C-series rotation curve (the right
panel of Figure~\ref{fig:rotcurve}). The overall rotation curves for
the F- and C-series models are  within 5 per cent at any given radius,
despite the relative
contributions from each component being appreciably different. The C-series rotation curves have
$f_D=0.81$ at $R=2.2R_d$.
As we shall see, the resonant dynamics are quite
different, as the phase-space gradient of the distribution function is
the controlling parameter of the resonant density (see
\citealt{sellwood14rev} for a review). In a cored halo, the flattening
of the central density profile creates a harmonic core, i.e. orbits at
small radii would execute simple harmonic motion in the absence of a
stellar disc. For this model series, the relatively low halo density near the centre of the simulation leads to a
well-known buckling instability (see \citealt{sellwood14rev} for a review). The differences between the fiducial
and variant halo models are discussed in section
\ref{subsec:haloinitials}. 

We construct rotating haloes consistent with the cumulative mass
distribution $M$ for specific angular momentum $j(M)$ in a spherical
shell as proposed by \citet{bullock01},
\begin{equation}
  j(M) \propto M^{1.3},
  \label{eq:jcum}
\end{equation}
normalised such that
\begin{equation}
  \lambda=\frac{\sum_i
    m_i\vec{r_i}\times\vec{v_i}}{\sqrt{2}M_{\rm vir}V_{\rm vir}R_{\rm
      vir}}=0.03.
  \label{eq:lambda}
\end{equation}
The value $\lambda=0.03$ is roughly the median value in the
distribution compiled from cosmological simulations by
\citet{bullock01}.  We realise a distribution that satisfies equations
(\ref{eq:jcum}) and (\ref{eq:lambda}) as follows.  First, we begin with a
phase space for the initially nonrotating spherical halo as described
in section \ref{subsubsec:fiducial}.  We randomly sample the phase
space of the initially nonrotating halo by mass as described in section
\ref{subsubsec:fiducial} and choose particles from the
distribution $M(r)$ by rejection.  Then, we change the direction of rotation for
orbits with $L_z<0$ from the probability distribution determined from
equation (\ref{eq:jcum}) by changing the sign of $L_z$ until the desired value of $\lambda$ is
obtained. For our $N=10^7$ haloes, this results
in a deviation from the desired $\lambda=0.03$ of less than 1 per cent.
Changing the sign of any component of the angular momentum of a particular orbit
remains a valid solution to the collisionless Boltzmann equation,
preserving the initial equilibrium of the system. Large values of
$\lambda$ may result in an unstable halo \citep{kuijken94}, but this
is not observed for the modest $\lambda$ selected here, unsurprising
given the results of \citet{barnes86} and \citet{weinberg91}. The initial
rotation curves for Fr and Cr (see Table~\ref{tab:simlist}) are
the F and C series respectively (Figure~\ref{fig:rotcurve}).

\subsubsection{Two tests of the dynamical mechanism}
\label{subsubsec:numericalvariants}

In addition to the halo model variants, we examine two additional
modifications to the fiducial model designed to test our physical
understanding of the dynamical processes. Such restricted simulations
do not have analogues in the real universe, but can illuminate specific processes.

In the first experiment, Ff in Table \ref{tab:simlist}, we exploit the separate
Poisson-equation solutions for the stellar disc and dark matter halo in EXP (see below for a
description of this N-body code) 
by allowing the
dark matter halo potential to self-consistently evolve while disallowing
changes to the stellar disc potential. This allows us to investigate
the adiabatic compression of the initially spherical dark matter halo
in response to the stellar disc potential. The results are compared to
that of the fiducial simulation in section \ref{subsubsec:haloorbit}
to isolate the effect of non-axisymmetric disc evolution on
the global halo properties.

In the second experiment, Fs in Table \ref{tab:simlist}, we artificially eliminate the
trapping that leads to the shadow bar by shuffling the
azimuth of halo particles in the fiducial simulation. The shuffling
interval for an individual orbit $i$ is chosen following a Poisson
distribution with an average value of $2P_{\phi,i}$ where $P_{\phi,i}\equiv
2\pi/\Omega_{\phi,i}$ is the azimuthal period for orbit $i$. This technique preserves the
radial structure of the halo and produces minimal disturbances to the
equilibrium. The results of this experiment are used to quantify the
effect of trapping on classical dynamical friction torque and to
corroborate our hypothesis of the shadow bar's effect on the angular
momentum transport between the disc and halo in section
\ref{subsec:effect}. 

We verify that the shuffling process effectively
eliminates trapping by attempting to detect a trapped component using
the methodology that will be described in \ref{subsec:methodology},
finding that $<$0.1 per cent of the orbits demonstrate a trapped
signal, consistent with uncertainties owing to resolution. Both models use the F-series
initial rotation curve (Figure~\ref{fig:rotcurve})

\begin{figure*}
  \centering 
  \includegraphics[width=6in]{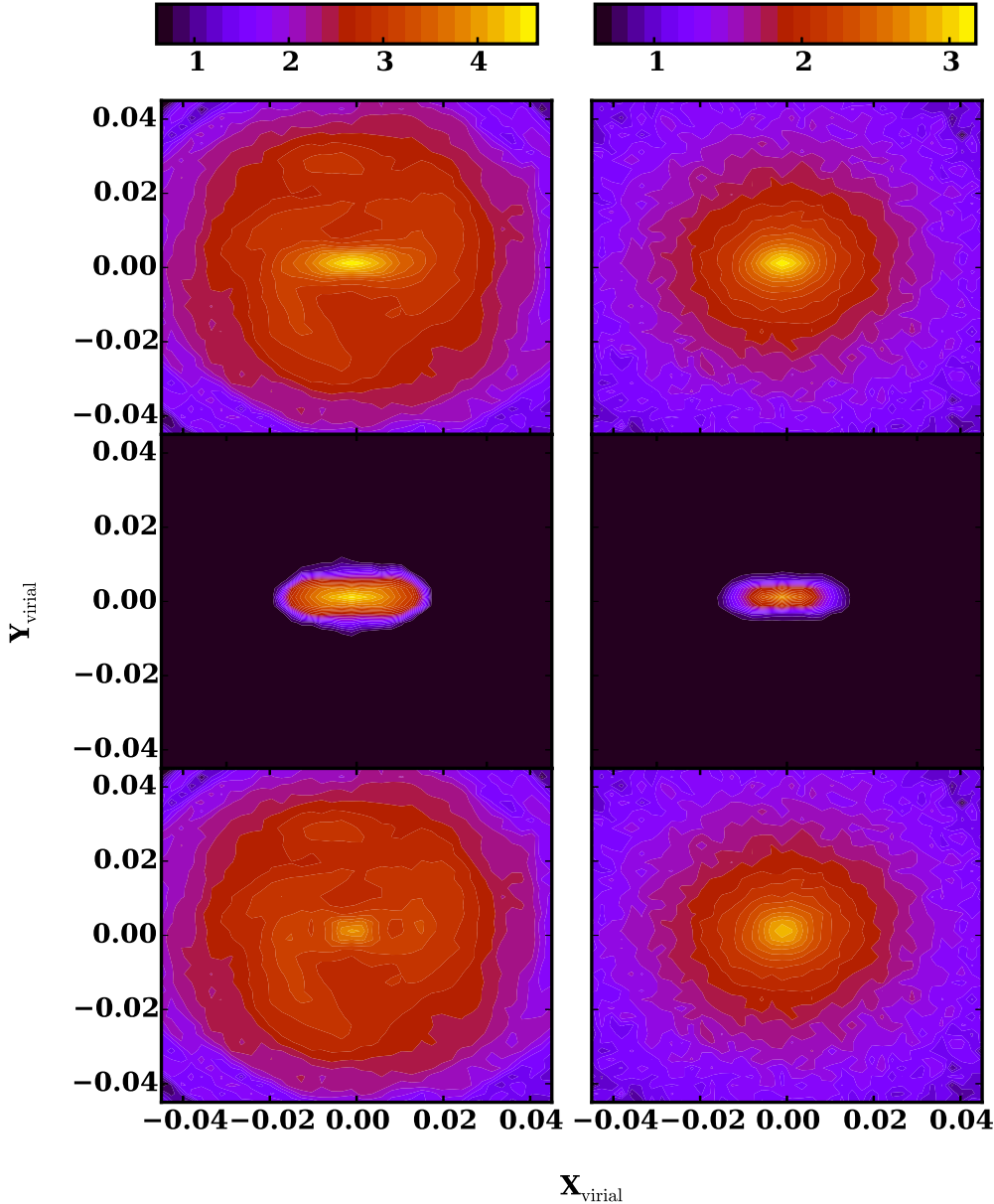} 
  \caption{The surface density of the fiducial simulation for the
   stellar disc (left panels) and the dark matter halo (right
    panels). The upper panels show the in-plane density ($|z|<0.003$,
    $\approx$ 1 kpc for MW-like scalings), the middle
    panels show the in-plane density of the trapped component (the
    stellar bar on the left and the shadow bar on the right), and the
    lower panels show the in-plane density of the untrapped component,
    at T=2.0 (4~Gyr for MW-like
    scalings). Positions are given in virial units, as in
    Fig.~\ref{fig:rotcurve}, discussed in section~\ref{subsec:methods}. \label{fig:shadowbar}} 
\end{figure*}

\subsection{N-body simulations} \label{subsec:integration} 

We perform the system evolution using EXP \citep{weinberg99}. This code
is designed to represent the gravitational field for multiple galaxy
components by a rapidly converging series of functions.  EXP is
advantageous for secular dynamics problems owing to its absence of
high-frequency noise and its high-efficiency relative to fully adaptive
Poisson solvers such as direct summation or tree gravity algorithms.

These series of functions are the eigenfunctions of the Laplacian.
For conic coordinate systems, the Laplacian is a special case of the
Sturm-Liouville (SL) equation,
\begin{equation}
  \frac{d}{dx}\left[p(x)\frac{d\Phi(x)}{dx}\right]-q(x)\Phi(x) =
  \lambda\omega(x)\Phi(x)
  \label{eq:SL}
\end{equation}
where $\lambda$ is a constant and $\omega(x)$ is a known function,
called either the density or weighting function.  These eigenfunctions
have many useful properties.  For systems with a finite domain, the
eigenvalues are real, discrete, and bounded from below; their
corresponding eigenfunctions are mutually orthogonal, and oscillate
more rapidly with increasing eigenvalue.  We will assume that the functions
are indexed in order of increasing eigenvalue. For example, the eigenfunctions of the Laplacian
over a periodic finite interval are sines and cosines; in cylindrical
coordinates, they are Bessel functions.  For a spherical geometry, the
eigenfunctions satisfy Poisson's equation such that
\begin{equation}
  \frac{1}{4\pi G}\int dr r^2 d^*_k(r)u_j(r) = \delta_{jk}
\end{equation}
where $d_k$ is the density and $u_j$ is the potential that together form a
biorthogonal pair. A variety of analytic biorthgonal pairs are
available (\citealt{cluttonbrock73}, \citealt{cluttonbrock77},
\citealt{hernquist92a}, \citealt{earn96}).  Here, we follow
\citet{weinberg99} and numerically solve equation (\ref{eq:SL}) to
obtain a basis whose lowest-order pair ($n=0$, $l=m=0$) matches the
equilibrium profile of the initial conditions. Higher-order terms then
represent deviations about this profile with successive higher spatial
frequency, which can account for the evolution of the system.

The overall power owing to discreteness noise scales as $1/N$, and we
may limit small-scale fluctuations by truncating the series,
effectively providing a low-pass spatial filter that removes
high-frequency noise from the discrete particle distribution
\citep{weinberg98}.  Although it is possible to derive an analogous
three-dimensional basis from the cylindrical Laplacian, the boundary
conditions are complex and difficult to match to the approximately
spherical domain of the halo.  After much trial and error, we found
that a empirical orthogonal function decomposition of a
high-dimensional spherical basis produces better results.
\citet{weinberg99} and HB05 describe this method in detail. The field
approach is also advantageous as it enables a direct comparison with
perturbation theory. However, we are aware of two significant
disadvantages of the field approach as well: 1) the technique cannot
be fully adaptive and limit high frequencies simultaneously; and 2) it
can be susceptible to $m=1$ (dipole) modes induced by unphysical
`sloshing' of the expansion centre against the centre-of-mass.  While
our simulations show nonzero $m=1$ power, we verify that the expansion
centre deviates from the centre-of-mass by less than $0.1z_0$ at all times. We
also note that $m=1$ modes are theoretically predicted
\citep{weinberg94} and appear in real galaxies
(e.g. \citealt{zaritsky13}) and hence do not necessarily indicate a numerical
fault. We will investigate these phenomena further in a future paper.

For this class of Poisson solver, the computational effort scales
linearly with particle number, making low-noise simulations with high
$N$ possible. A number of basis terms are kept to allow for astronomically realistic
perturbations to the equilibrium profile while minimising the linear
least-squares fit to a Monte Carlo realisation of the initial particle
positions. We retain terms in the halo basis up to $l=m=6$, and a
maximum radial index of $n=20$ corresponding to the largest
eigenvalues for the spherical harmonics of the halo. For the disc, we
select $m=6$, $n=12$ for the empirical orthogonal functions. This
basis selection allows for variations of $0.1 R_d$ (300 pc for a
MW-sized galaxy) in the vicinity of the bar, decreasing to $5\times
10^{-5}$ (15 pc for a MW-sized galaxy) near the centre.  This solver naturally suppresses the small scales that may lead to anomolous diffusion, which can lead to unphysically rapid evolution, without adding spatial resolution on the scales of interest\footnote{Note that this method does not remove all the
  relaxation from fluctuations in the basis; this method will only
  remove relaxation on resolution scales smaller than those probed by
  the basis.}. While possible, we do not recondition the basis on the
evolving equilibrium during simulations, relying instead on the
initial Poisson variance to allow for sufficient degrees of freedom.

Particles are advanced using a leapfrog integrator with an adaptive
time step algorithm. Briefly, we begin by partitioning phase space $s$
ways such that each partition contains $n_j$ particles that require a
time step $\delta t=2^{-j}\delta T$, where $\delta T$ is the largest
time step and $j=0,\ldots,s-1$.  The time step for $j=s-1$ corresponds
to a single time-step simulation.  Since the total cost of a time step
is proportional to the number of force evaluations, the speed up
factor is given by: ${\cal S} = \sum_{j=0}^{s-1}n_j/\sum_{j=0}^{s-1}
n_j 2^{-j}$. For a $c=15$ NFW dark-matter profile with $N=10^7$
multimass particles as described in section~\ref{subsubsec:fiducial},
we find that $s=8$ and ${\cal S}\approx 30$, an enormous speed up!
Forces in our algorithm depend on the basis coefficients and the leap
frog algorithm requires linear interpolation of these coefficients to
maintain second-order error accuracy per step.  The expansion
coefficients are partially accumulated and interpolated at each level
$j$ to preserve continuity as required by the numerical integration
scheme for finer time steps.  This interpolation and the bookkeeping
required for the successive bisection of the time interval is
straightforward.  The symmetric structure of the leap-frog algorithm
allows the coefficient values to be interpolated rather than
extrapolated.  This algorithm will be discussed in more detail in a
forthcoming paper.

To motivate our choice of timesteps, we first define a
characteristic force scale as $d_i\equiv
\Phi_i/|\partial\Phi_i/\partial\mathbf{x}_i|$ where $\Phi_i$ is the
current scalar potential of the particle. The time step level, $j$,
for each orbit is then assigned by the minimum of three criteria: (1)
$v_i/a_i$, the characteristic force timescale, (2) $d_i/v_i$, the
characteristic work timescale, and (3) $\sqrt{d_i/a_i}$, the
characteristic escape timescale. The quantity $v_i$ is the current
scalar velocity of the particle and $a_i$ is the current scalar
acceleration of the particle. The criteria have prefactors
$\epsilon_f,~\epsilon_w,~\epsilon_e$, respectively, which can be tuned
to achieve the desired time resolution. Using the initial distribution
of particles in the fiducial model, we tune the ratios of the
prefactors based on the mean of the ratio of the calculated timescales
for all the disc particles. We find that $\epsilon_f/\epsilon_w\approx
10$, and $\epsilon_w/\epsilon_e\approx 2.5$.

\citet{weinberg07a} use numerical perturbation theory to determine a
scaling for particle timestep criteria, finding that 1/100 of the
period of oscillation appropriately recovers the resonant
dynamics. \citet{weinberg07a} defined particle number and time-step
criteria to help ensure that the dynamics in the vicinity of the
homoclinic trajectory are accurate for a very slowly evolving system.
The formal requirement of a slowly evolving system is implicit in
the time ordering that enables the perturbative analytic estimates
that underpin secular evolution.  In this standard formulation of
secular evolution (e.g. dynamical friction), changes in otherwise
conserved actions occur during their homoclinic passage. For strong
(typically low-order) resonances, these regions are large and the
criteria are easily satisfied. For weaker (typically higher-order)
resonances, large particle numbers are required. Conversely, for more
rapidly evolving systems, a smaller particle number may suffice. We
choose to err on the side prudence by satisfying the slow-evolution
criteria. Using the fiducial
model, we find that at a scale height, $r=0.001$, the appropriate
timestep is $h\le1.8\times10^{-4}$, similar to the timestep chosen in
HB05, $h=2\times10^{-4}$. To comply with the findings of
\citet{weinberg07a}, we set the smallest timestep to be
$h=1.25\times10^{-4}$. We then tune the prefactors such that nearly
all the disc particles ($>99.5$ per cent) as well as halo particles in
the vicinity of the disc reside in this level, while allowing other
halo particles to occupy larger timestep levels to take advantage of
the computational speedup. We allow for four multistep levels ($s=5$) so
that the maximum timestep is 16 times larger than the minimum time
step ($h_{\rm max}=2\times10^{-3}$). We note that while many halo
particles at large radii do not even require $h=2\times10^{-3}$ for
1/100$^{\rm th}$ of an oscillation, we truncate the multistep levels
to match the output frequency, $\delta T=0.002$. While this decreases
the speed up factor, we find that with $s=5$, $S\approx5$, still a
considerable speed up.

\section{Results from the fiducial simulation} 
\label{sec:results}

This section characterises the fiducial simulation.  We begin with a
description of the structure and long-term evolution of the bar
profile and pattern speed in section \ref{subsec:orbitstat}. We then
motivate a new time-dependent tool for characterising orbits trapped
into resonance and apply this tool in section
\ref{subsubsec:discorbits} and \ref{subsubsec:haloorbit} to the disc
and halo populations, respectively.  We show that the evolution of a
stellar bar in the presence of a dark-matter halo traps a large
fraction of dark matter orbits in the vicinity of the bar, resulting
in a shadow bar. The implications and dynamical consequences of
this trapping on the evolution of the galaxy are presented in section
\ref{subsec:dynquan}. This is followed by a discussion of the
observational consequences of the shadow bar in section
\ref{subsec:effect}.

\subsection{Formation and evolution of the bar} 
\label{subsec:orbitstat}

Figure~\ref{fig:rotcurve_evolved} shows the rotation curves determined by rank
ordering the force-derived circular velocities (as in
Figure~\ref{fig:rotcurve}, $v_c^2=rF_r$, calculated separately for each component) of particles in a narrow annulus at
each radius $R$ and selecting quantile values from $p=0.2-0.8$.
We display the resulting curves for two characteristic times: immediately after initial bar formation
($T=0.4$) while the bar is still rapidly growing, and well-after bar
formation ($T=2.0$).  We display the range of circular velocities
spanned by the $p=0.2-0.8$ quantile values to demonstrate the
non-axisymmetric nature of the circular velocity field (particularly for radii $R<R_{\rm bar}\approx0.01$), i.e. the minor
axis of the bar will have appreciably higher circular velocity than
the major axis of the bar. The bounding $p=0.2,0.8$ quantiles are chosen to be
representative of the approximate values along the major and minor
axes, respectively. The bulk of our analyses focus on
the evolution after a discernible bar feature has formed,
i.e. $T>0.4$. The rotation curve continues to evolve after bar
formation, as evidenced by the comparison between the 
panels of Figure  \ref{fig:rotcurve_evolved}. In particular, the disc
contribution responds strongly to the formation and continued
evolution of the bar, though the halo contribution also changes by up
to 25 per cent at small radii ($R<R_{\rm bar}$).

Initially, the ratio of dark matter mass to stellar mass inside of one disc scale
length is $M_{\rm dm}/M_{\star}(R<0.01)=0.757$.  Although the
gravitational potential in the disc plane is dominated by the stellar
contribution, the dark matter provides an important channel for
secular evolution, changing its phase-space distribution by trapping
has important dynamical consequences that we will outline
below\footnote{At late times, $T=2.0$, $M_{\rm dm}/M_{\star}(R<0.01)=0.595$,
  owing to an increase of stellar mass at small radii--a consequence
  of bar formation.}.

In Figure~\ref{fig:shadowbar}, we plot the surface density of the disc
and halo at a late time ($T=2.0$), well after the bar has formed. An
in-plane density slice of the halo reveals an approximately elliptical
distribution aligned with the stellar bar, up to a maximum ellipticity
of $e=0.3$ in the fiducial model. Because the disc bar dominates the
gravitational potential at these radii, by itself, this is not
particularly surprising (e.g. \citealt{colin06};
\citealt{athanassoula07}; \citealt{debattista08}). However, as we
shall see in section \ref{subsubsec:haloorbit}, a large fraction of
this dark matter is dynamically trapped in the bar potential; i.e.
it now part of the bar! The dynamical nature of this halo feature and
its evolutionary implications has not been identified
previously. Additionally, a three-dimensional analysis of the disc
shows that the initially spherical halo has, at late times, been
flattened by the presence of the disc potential particularly at
$R<0.01$ (see section \ref{subsubsec:haloorbit} for additional
discussion).

\begin{figure} 
  \centering 
  \includegraphics[width=3.4 in]{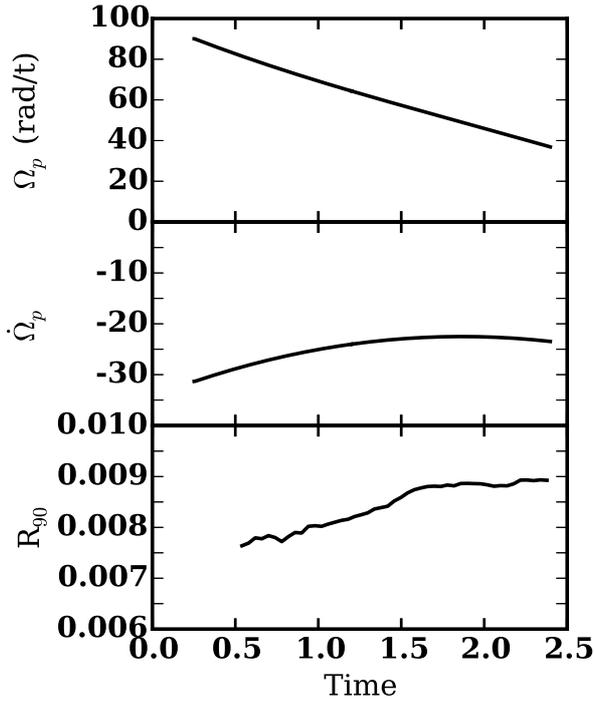} 
  \caption{Upper panel: the pattern speed of the stellar bar as a function of time
    for the fiducial simulation. Middle panel: the rate of change in the
    pattern speed as a function of time. Bottom panel: the
    characteristic radius of the bar that encloses 90 per cent of the
    trapped stellar orbits ($R_{90}$) determined by $k$-means
    analysis. The bar slows nearly linearly at all times while
    the bar lengthens. Radius and time are given in virial units. \label{fig:patternspeed}}
\end{figure}

We first describe the stellar bar by fitting ellipses to the projected
disc surface density at each time. The bar changes appreciably with
time, both elongating and slowing over the course of the
simulation. In the upper panel of Figure~\ref{fig:patternspeed}, we plot
the pattern speed of the bar, $\Omega_p$ as a function of time. The
pattern speed monotonically decreases with time, nearly linearly as
$\dot{\Omega}_p$ changes only slightly (middle panel).

This global, phenomenological view of bar evolution does not reveal
the underlying dynamical details and their implications for the
long-term evolution of galaxies.  Advancing beyond the traditional
ellipse and Fourier determinations of the bar is possible in N-body
simulations, albeit difficult. The centrepiece of our analysis below
is the ability to accurately examine the dynamical characteristics of
individual orbits to construct a more nuanced understanding of
bar-disc-halo dynamics.

\subsection{Orbit taxonomy} \label{subsec:methodology}

\begin{figure*} 
  \centering 
  \includegraphics[width=6.5 in]{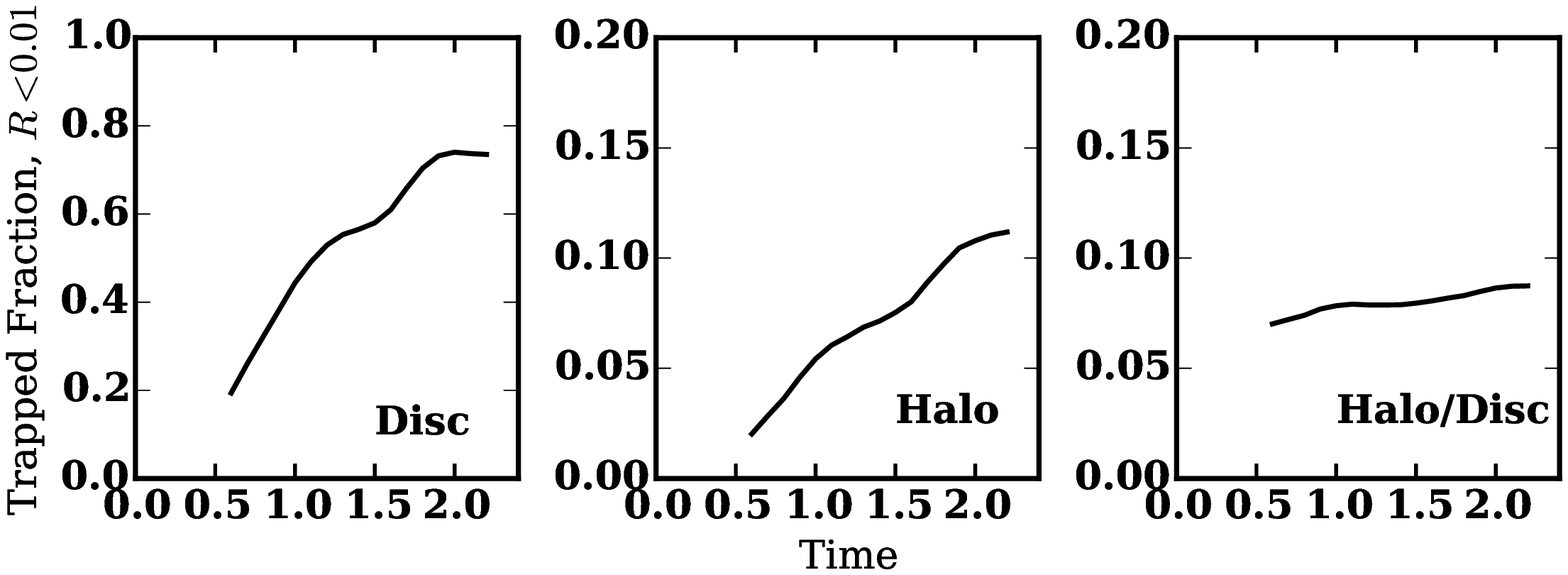} 
  \caption{Trapped populations by fraction of mass interior to
    $R=0.01$ (the initial scale length), as determined by $k$-means power, for the fiducial
    simulation. Left panel: disc trapping as a function of time.
    Middle panel: halo trapping as a function of time.  The
    trapped halo fraction (the shadow bar) monotonically increases at
    all times. Right
    panel: The ratio of the shadow bar to stellar bar (by total mass) as a function of
    time.  The halo is trapping at a
    faster rate than the disc, owing to the consistent increase in the
    reservoir of untrapped matter as the halo contracts in response to the
    bar growth. Time is given in virial units. \label{fig:disctrap}}
\end{figure*}

Orbits mediate the transfer of angular momentum between large-scale
components (e.g. the disc and halo).  For systems dominated by regular
orbits for which the derivatives of the potential can be computed exactly
(i.e. separable and regular systems), Fourier transformations of
particle coordinates can determine the orthogonal frequencies of
orbits in cylindrical coordinates ($\Omega_r,~\Omega_\phi,~\Omega_z$)
and these frequencies allow for rapid and unambiguous determination of
orbits trapped into potential features. The most obvious of these
features are resonances, commensurabilities of frequencies with the
pattern speed of the bar that satisfy the relation
\begin{equation}
  l_r\Omega_r + l_\phi \Omega_\phi = m\Omega_p.
  \label{eq:resonances}
\end{equation}
Previous studies have analysed N-body simulations by fixing or
`freezing' the gravitational potential after a period of
self-consistent evolution and characterise the time-independent
orbital structure (e.g. \citealt{binney82}, \citealt{martinez06},
\citealt{saha12}). If the rate of bar evolution is sufficiently small
compared to the time required to cross the resonance zone ($\dot{\Omega}_p=\mathcal{O}(\epsilon)$ and $\dot{I}_{\rm
  bar}=\mathcal{O}(\epsilon)$ as $\epsilon\rightarrow0$, where
$\dot{I}_{\rm bar}$ is the time derivative of the moment of inertia of the bar), the dynamics
will be approximately described by secular theory
(e.g. \citealt{tremaine84}, \citealt{weinberg85},
\citealt{hernquist92}). However, realistic potentials are continuously
evolving, so one must attempt a finite-time analysis of secular
processes to investigate the nonlinear effects.  In other words, one
must devise techniques that characterise orbits while the bar pattern
speed and geometry change.  To this end, we have developed a $k$-means
classifier for orbits, which quickly and accurately decomposes particles
into orbital components classified by their
commensurabilities. Briefly, the $k$-means technique \citep{lloyd82}
iteratively sorts a collection of points into $k$ clusters given a
distance metric. For our application, we use a time series in the $r$
and $\theta$ (azimuthal angle) positions of apsides (radial turning
points, $r_{\rm max}$) for a given orbit to determine its membership in
the bar. As a first application of the method in this work, we limit
our analysis to $k=2$.

In a frame corotating with the bar, the apsides of orbits trapped by
the bar do not precess through $2\pi$.  That is, the azimuthal
position of outer turning points for trapped orbits oscillate about
some fixed position angle in the bar frame of reference.  We exploit
these slowly changing apse positions for trapped orbits to identify
coherent bar-supporting orbits over a finite time period.
Specifically, we define $\tpara$ to be the absolute value of the
difference in azimuthal angle between the outer turning point of a
particular orbit and the bar position angle.  Let $\tparaL_{20}$ be the
average value of $\tpara$ in $20$ azimuthal periods for a given orbit. We
use the $k$-means technique to identify orbits with
$\tparaL_{20}<\pi/8$.  Orbits that are periodic and aligned with the
bar in its corotating frame have $\tparaL_{20}=0$ while those with no
correlation have $\tparaL_{20}=\pi/4$.  Additionally, relevant
resonances trap orbital apsides perpendicular to the bar.  In this
case, we define an analogous $\tperp$.  In practice, this technique
allows for rapid determinations of bar membership over time windows
that are too brief for frequency determinations using Fourier
techniques.  In many cases, the bar evolution is so rapid that an
analysis in a frozen potential would not be relevant. 

After testing a range of period windows, the choice of 20 azimuthal
periods was chosen as a balance between minimizing sampling noise
while retaining temporal resolution of the evolutionary time scale. Our method is
applicable to all but the most extremely centrally confined particles whose
azimuthal periods are too small to satisfy an approximate Nyquist
criterion.  For the simulations here, this critical azimuthal period
is $T_\theta = 0.004$ and affects approximately 0.1 per cent of the
orbits. As a simple verification test, we analysed the 
fixed disc potential simulation (i.e. non-axisymmetric disc evolution is
disallowed, Ff in Table~\ref{tab:simlist}), finding a bar signal of
$<$0.1 per cent, demonstrating that the method is very robust against
false positives.

For this paper, we will abbreviate our discussion by assuming that all
orbits trapped by the bar potential are 2:1 orbits--those that
exhibit two radial periods in every azimuthal period, identified by
$k=2$.  In reality, the primary resonance bifurcates into 4:1 and 2:1
orbits at high bar strength near the ends of the bar
\citep{athanassoula92}.  An in-depth discussion of the $k$-means
identifier and its application to orbital family determination will be
presented in a later paper.

\subsubsection{Disc orbits} \label{subsubsec:discorbits}

Since our understanding of secular evolution is informed by
perturbation theory, we describe the orbital behaviour in the disc by
conserved quantities; here we choose energy and the in-plane angular
momentum, $E$ and $L_z$. At radii larger than the corotation radius (hereafter,
CR)\footnote{The CR resonance has $m=l_\phi=2,~l_r=0$ in the notation of
  equation \ref{eq:resonances}.  The CR radius is $\approx0.02$ in
  the fiducial simulation at $T=2.0$.}, orbits have low ellipticity
($e<0.1$).  Inside of CR, the ellipticity increases owing to the bar.
The bar-parenting orbit family (${\bf x}_1$ in the notation of
\citealt{contopoulos80}) is an orbit family whose outer turning points align.  These arise from the inner Lindblad resonance
(ILR)\footnote{The ILR resonance has $m=l_\phi=2,~l_r=-1$ in the notation of
  equation \ref{eq:resonances}} in perturbation theory. Such orbits
are easily identified with the $k$-means identifier.

As a first application of the $k$-means method, we identify the radius
that encloses 90 per cent  of the bar-trapped disc orbits as the
characteristic bar radius, $R_{90}$. The bottom panel of Figure
\ref{fig:patternspeed} shows that the radius of trapped orbits
increases monotonically with time, corroborating previous works
indicating the lengthening of bars over time.  The trapped fraction of
disc orbits satisfying $R<0.01$ is an explicit measure of the bar mass
and, therefore, the bar potential (see the left-hand panel of Figure
\ref{fig:disctrap}). The fiducial simulation exhibits a monotonic
increase in the trapped fraction with time. Further, the growth of the
bar can be roughly divided into three phases: formation, fast growth,
and slow growth. The first, formation ($T<0.5$ in
Figure~\ref{fig:disctrap}), is not studied in this work. We instead begin our study of orbits at times after
which the bar can be detected through ellipse fitting, the fast growth
phase\footnote{The trend at $0.5<T<1.0$ cannot be extrapolated to
  T=0.0, suggesting that the growth rate at $0.0<T<0.5$ is more rapid
  than in the range $0.5<T<1.0$. However, this cannot be determined
  through the $k$-means technique due to time resolution limits.}. In this phase ($0.5<T<1.2$), the bar
rapidly gains mass and significant $m=2,4,6$ patterns are observed
beyond $R_d$. In the third phase ($T>1.2$), the bar continues to add mass (albeit at a
diminished rate), but
the disc is relatively devoid of other $m\ne 2$ features.

\begin{figure}
  \centering 
  \includegraphics[width=3.5in]{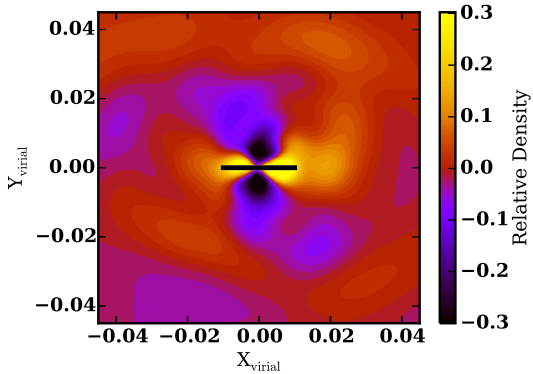} 
  \caption{The halo wake ($(m>0)
    + (m=0, n\ne0)$, filled
    colour) with the bar position and length indicated as the thick black
    line.  The wake lags behind the bar (rotating
    clockwise) at $R>0.5R_d$, but is generally aligned with the
    bar within $0.5R_d$, where we observe significant trapping (see
    Fig.~\ref{fig:disctrap}). Note the resemblance to the untrapped
    component of the disc in
    Fig.~\ref{fig:shadowbar}. Positions are given in virial units, as in Fig.~\ref{fig:shadowbar}.\label{fig:halowake}}
\end{figure}

\subsubsection{Halo orbits}  \label{subsubsec:haloorbit} 

The structure of the dark matter distribution in the stellar disc has
recently been debated, with particular discussion regarding the
creation of a co-rotating `dark disc' \citep{read08,bruch09,read09,purcell09,pillepich14}. This
dark disc is formed by the accretion of satellites that settle into a
rotating structure through dynamical friction, affecting the local
dark matter density by contributing up to 60 per cent in some
simulations \citep{ling10}, but very little in others
\citep{pillepich14}. Additionally, the halo is shown to be prolate at
small radii in the presence of a stellar bar, referred to as the `halo
bar' or `dark matter bar' \citep{colin06,athanassoula07,athanassoula13,saha13}, though
\citet{debattista08} demonstrated that if the disc were evaporated,
the prolate inner region of the halo would recover its initially
spherical shape.

In our fiducial simulation, the dark-matter halo becomes oblate near
the disc at late times. We also refer to this as the `dark disc',
while noting the difference in formation mechanism from previous
authors. This dark disc is a necessary consequence of the gravitational
  potential of the disc instead of depending on accretion. We,
  therefore, expect this dark matter feature to be a generic
  consequence of stellar discs. At $T=2.0$, long after the bar has formed, and at
$R=0.03$, approximately the radius where
the disc and halo contributions to the rotation curves are equal, a
determination of the ellipsoidal shape of the halo following the
method of \citet{allgood06}
finds $S\equiv c/a=0.45$ (as well as $Q\equiv b/a=0.8$), where
$a$, $b$, and $c$ are the lengths of the major, intermediate, and
minor axes, respectively. While some of this flattening is attributable
to adiabatic compression of the halo in response to the presence of the stellar disc, a test
simulation that allows the halo to achieve equilibrium while holding
the initial disc potential fixed demonstrates that adiabatic
compression alone, which gives $S=0.6$ and $Q>0.99$, 
cannot account for their deviation from the initially spherical distribution. 
In the plane of the disc, the effect is even
more pronounced; at $R=0.03$ for all $T>1.0$, the azimuthal average of the density in
the disc plane is $>$30 per cent larger than the initially spherical
distribution\footnote{The density along the polar axis at all times is nearly
  equal to the initial distribution at $R>0.01$. At $R<0.01$, the
 potential of the disc causes spherical adiabatic
 contraction such that the halo density along the polar axis is also
 enhanced relative to the initial distribution.}.

Examining the evolution of the halo flattening in the fiducial
simulation reveals that the initial response of the halo to the
presence of the disc is complete by $T=0.3$, in the midst of bar
formation; at $T=0.3$ and $R=0.03$ in the fully self-consistent
simulation we find that $S=0.6$ and $Q=0.95$.
The similarity of these values to that
of the settled halo in the fixed disc potential test at late times
suggests that the evolution of the disc is responsible for the continued
compression and the move to triaxiality. Furthermore, the continued evolution
of the halo shape suggests that simply including adiabatic compression
in halo models is insufficient to recover the dynamics
of the stellar disc and halo.

The distribution of $\cos\beta\equiv
L_z/L$ also illuminates the flattening of the halo\footnote{The
  definition of $\cos\beta$ means that an orbit in the disc plane will
  have $\cos\beta=1$, such that $\beta=0^\circ$. Throughout the work, we will make use of the limit
  $\cos\beta>0.9$, which corresponds to $\beta<26^\circ$, to define halo
  orbits at low inclination relative to the disc angular momentum plane.}. A
perfectly spherical halo will exhibit uniform coverage in $\cos\beta$
owing to the random orientations of the orbital planes. However, at
high binding energy (small radii), where the disc dominates the
potential, the halo shows enhanced populations at
$\cos\beta\approx1$. Taken together, the continued flattening of the halo and the distribution
of $\cos\beta$ suggest that the continued
deposition of angular momentum into the halo owing to secular torques
from the bar causes the formation of the dark disc in our simulations.

Clearly, the perturbative effect of the stellar bar on the halo cannot be
ignored.  To investigate the secular response of the halo orbits to
the bar, we apply the $k$-means analysis to these orbits as well
and find a class of 2:1 orbits that
respond similarly to the bar as their disc analogues. In other words,
the presence of the stellar disc traps initially spherical halo orbits
into not only planar orbits, but into bar-supporting orbits that we
call the \emph{shadow bar}. We emphasise that this result is
qualitatively different from those of \citet{debattista08}, whose
orbits are necessarily untrapped if they retain their box-like nature
relative to the triaxial halo 
upon evaporation of the disc. We cannot comment on the likelihood of a
similar box orbit-retaining outcome for other studies that point out
a bar-like feature in the halo \citep{colin06,athanassoula07,athanassoula13}.

The growth rate of trapped halo orbits also follows that of the
stellar bar, albeit with a smaller fraction of the halo mass available
for trapping at $R<0.01$ (the middle panel of Figure
\ref{fig:disctrap}). The continued growth of the trapped fraction
during the simulation corroborates that halo trapping is secular and
not merely a result of the initial settling of the halo in response to
the stellar disc. In the rightmost panel of Figure~\ref{fig:disctrap},
we plot the ratio of dark matter trapped mass to stellar trapped mass
(the total for the simulation rather than that scaled to the mass interior to
$R_d$). This also rises monotonically throughout the simulation, owing
to both the presence of a larger reservoir of planar material as the
dark disc grows with time and the increasing strength of the bar
potential with time. The result is that the shadow bar grows at a
faster rate than the stellar bar.

Overall, the trapped halo orbits exhibit the same behaviour as their
trapped disc counterparts: bar-supporting orbits are found at larger
radii through time.  The trapped orbits in the halo are 5-10 per cent 
of the trapped mass of disc orbits, and 3-5 per cent of the total
trapped angular momentum.  That is, the shadow bar is not a large
contributor to the mass of the visible stellar bar.  However, the
existence of shadow bar modifies the rate and location of angular
momentum exchange between the bar and halo.  Simple perturbation
theory calculations following the model presented in
\citet{weinberg85} demonstrate that the halo orbits that accept $L_z$
from the disc are preferentially located near $\cos\beta=1$ and these
orbits are the ones that are efficiently trapped into the shadow
bar. If we bin the halo orbits by $\cos\beta$, we find that at
$T=2.0$, approximately 45 per cent of the orbits satisfying
$\cos\beta>0.9$ in the annulus $0.003<R<0.01$ are trapped\footnote{The lower bound, $R=0.003$, is selected to
  avoid the inner region of the halo where the $z$-dimension
  oscillations of the orbit, even if the orbit has net rotation
  (nonzero $L_z$), confuse the calculation of $\cos\beta$.}, compared to
12 per cent for all dark matter at $R<0.01$ (see
Figure~\ref{fig:disctrap}). Preferential trapping of orbits with $\cos\beta>0.9$ suggests that the shadow bar will decrease the angular momentum
transport from the bar to the halo relative to that predicted by
dynamical friction theory.  We will estimate the magnitude of these
effects in section \ref{subsec:effect}. In
  the case of a dark disc formed through satellite accretion, whose
  distribution would be concentrated near $\cos\beta=1$ rather
  than evenly distributed in $\cos\beta$, we would expect to observe
  enhanced trapping relative to what is presented here.

Moreover, the overall contribution of the dark matter wake has a very
long reach.  To see additional effects of the shadow bar on the halo
at large scales, we plot the in-plane wake (as in
\citealt{weinberg07a}) at $T=2.0$ in Figure~\ref{fig:halowake}. Here,
we define the halo wake as the response of the \emph{untrapped} halo
  to the non-axisymmetric bar. We find
the halo wake by removing all the trapped orbits as found by the
$k$-means analysis and comparing the resulting distribution to the
lowest order axisymmetric density profile.  To accomplish this, we
constructed a new basis using only the untrapped orbits (as described in
section \ref{subsec:integration}) and then removed the lowest order
($m=0,~n=0$) component to find the wake\footnote{Note that in addition
  to the non-axisymmetric $m\ne0$ components, we include $m=0,~n\ne0$
  terms in the wake calculation.}. Despite removing the trapped
orbits, we observe a strong quadrupole feature that follows the
stellar bar (along the $x$-axis) and extends to the end of the stellar
bar (the extent of the stellar bar is shown as the black line). Beyond
the extent of the bar, an $m=2$ pattern is observed, a direct result
of the secular evolution detailed in \citet{weinberg85}.  If the disc
scale length of the simulation is scaled to that of the
MW, this wake extends out to the
Solar circle at approximately the 10 per cent level. This scaling of the simulation may not be the
  most appropriate due to the apparent disagreement between the bar
  length-scale length relationship in our simulation and the MW. If the
  simulation is instead scaled to the bar length of \citet{wegg15},
  4.6 kpc, the density enhancement from the wake may be as large as 25
per cent!

\subsection{Dynamical properties of orbits}  \label{subsec:dynquan}

\begin{figure}
  \centering \includegraphics[width=3.5in]{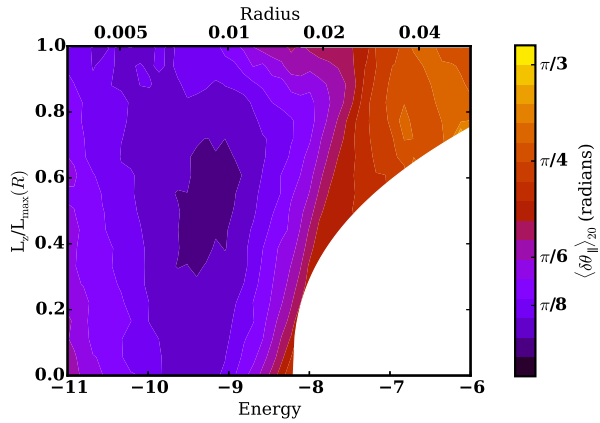} 
  \caption{The distribution of $\tparaL_{20}$ as a function of energy,
    $E$, vs. scaled angular momentum, $L_{\rm z}/L{\rm max}(R)$, at $T=2.0$ for the stellar disc.
    At each radius, $L_{\rm z}$ and $E$ are calculated for a circular
    planar orbit and used to determine the mapping between energy and
    radius, as well as the maximum $L_{\rm z}$ at a given radius.  The
    quantity $L_{\rm z}/L_{\rm max}(R)$ is zero for a radial orbit
    and unity for a circular orbit.  Nearly all disc orbits are prograde,
    $L_{\rm z}/L_{\rm max}(R)>0$.  The colours denote the average
    apse position relative to the position angle of the
    bar. The region in $E$-$L_{\rm z}$ space with
    insufficient density to obtain a reliable estimate is 
    white.  The disc is largely comprised of circular orbits at
    $R>0.02$. The dark blue region indicates the stellar
    bar. Radius and energy are given in virial units, see
    Section~\ref{subsec:methods}. \label{fig:betastellar}}
\end{figure}

To understand the dynamics of the trapped halo orbits, we investigate
the distribution of conserved quantities in the various orbit
populations.  We characterise the orbits by energy $E$, an isolating
integral for all regular orbits, and $L_z/L_{\rm max}(R)$, where we
normalise the angular momentum perpendicular to the disc, $L_{\rm z}$,
by the angular momentum of a circular orbit with the same energy,
$L_{\rm max}$\footnote{The quantity, $L_z/L_{\rm max}$, is equivalent
  to $\left(\cos\beta\right)\kappa$, where $\cos\beta\equiv L_z/L$ and
  $\kappa\equiv L/L_{\rm max}$, as defined in the literature
  (e.g. \citealt{tremaine84}).}.  Figure~\ref{fig:betastellar} shows
the 
distribution of $\tparaL_{20}$ for stellar orbits from the fiducial
simulation at $T=2.0$.  Orbits with $\tparaL_{20}<\pi/8$ reliably
indicate bar-trapped orbits.  As expected, the strongest signal comes
from mildly elliptical orbits with $L_z/L_{\rm max}(R)=0.5$ near the
end of the bar with $E=-9.5$ ($R=0.01$).  Many of the orbits within a
characteristic radius of the bar are trapped (exhibiting
$\tparaL_{20}<\pi/8$), though Figure~\ref{fig:disctrap} indicates that not
all orbits within $R=0.01$ are trapped.

\begin{figure}
  \centering 
  \includegraphics[width=3.5in]{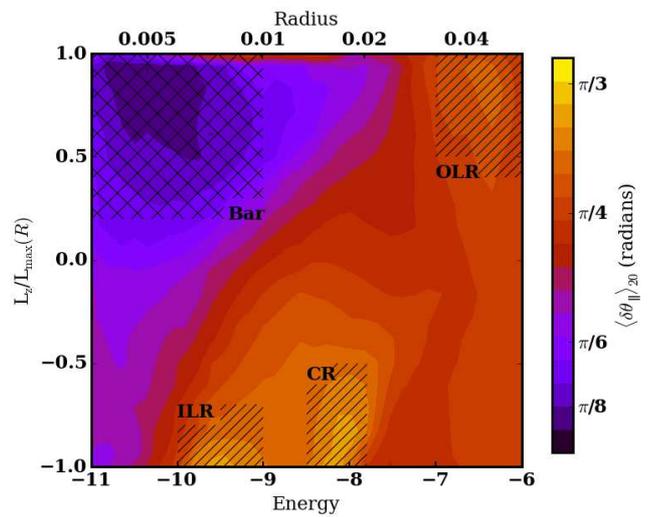} 
  \caption{As in Fig. \protect{\ref{fig:betastellar}} but for the dark
    matter halo.  Retrograde orbits are now present and the $L_{\rm
      z}/L_{\rm max}(R)$ axis now runs from -1 to 1. The dark blue
    region in the upper left, with an x-hatched region overlaid, indicates orbits trapped into the shadow
    bar, which occupies the same region in $E-L_{\rm z}$ space as the
    stellar bar (compare to Fig.~\ref{fig:betastellar}). Additional
    features from the $k$-means analysis are also seen, including two
    retrograde populations at $E=-9.5$ and $E=-8.2$ that are
    transiently moving through the ILR and the CR, respectively. A
    prograde population is observed at OLR, $E=-6.5$. These
    populations correspond to the wake overdensities in
    Fig.~\ref{fig:halowake}. As in Fig. \protect{\ref{fig:betastellar}}, radius and energy are given in virial units. \label{fig:betahalo}}
\end{figure}

A similar exercise performed for the halo (Fig.~\ref{fig:betahalo})
reveals that trapped halo orbits occupy the same region in $E-L_{\rm
  z}$ parameter space as the trapped disc orbits.  Because the
fiducial halo is initially isotropic the quantity $L_z/L_{\rm
  z,max}(R)$ can have values from -1 to 1.  The shadow bar reveals itself
by the prominent $\tparaL_{20}<\pi/8$ feature in the upper left corner
of Figure~\ref{fig:betahalo} at $E<-9$ ($R<0.01$).  This reinforces our
finding that both dark-matter and stellar orbits are trapped by their
mutual gravitational potential, and that the disc and the halo cannot
be treated as distinct dynamical components.

In addition to the shadow bar, Figure \ref{fig:betahalo} reveals two
planar retrograde families exhibiting $\tparaL_{20}>\pi/3$ (or
equivalently $\tperpL_{20}<\pi/6$), which correlate with overdensities
in Figure  \ref{fig:halowake} and are associated with the ILR and CR
resonances. Similarly, we observe the outer Lindblad resonance (OLR,
$m=l_\phi=2,~l_r=1$ in the notation of equation \ref{eq:resonances})
as a prograde planar family at $E=-6.5$ ($R=0.04$), observable
as an overdense region with $\tparaL_{20}>\pi/3$ in
Figure~\ref{fig:halowake}. The approximate position of the resonances
were located by using the monopole potential field of the disc and
halo from the basis so that equation \ref{eq:resonances} could be
solved numerically as a function of $E$ and $\kappa$ when combined
with the instantaneous pattern speed of the bar (see
Figure~\ref{fig:patternspeed})\footnote{Owing to the choice of a
  spherically symmetric system for the resonance determination, the
  tilt of the rotation plane, and thus our approximation
  for the location of resonances, does not depend on $\cos\beta$.}. 
We further confirmed the resonant nature of orbits in these $E-\kappa$ regions
by examining individual orbits.

To summarise the previous sections, the initially nearly isotropic
dark matter halo evolves significantly in the presence of an evolving
disc. The effects are three-fold: (1) the trapping of dark matter
orbits into the shadow bar; these orbits subsequently behave just like
stellar bar orbits; (2) the dark disc, a response to the presence of
the stellar disc as well as secular torques from the stellar and
shadow bar; these orbits do not resemble the stellar disc per
se, but more importantly do not resemble initial halo orbits; and (3)
the dark matter wake, created by the influx of $L_{\rm z}$ from the
stellar disc.

\subsection{Observational consequences of the shadow bar} \label{subsec:effect} 

Although the trapped dark-matter mass fraction is small, even a small change in
$L_z$ for halo 2:1 orbits affects both the structure of the dark
matter halo and the evolution of the stellar disc in our simulations,
and these dynamics are reflected in observable properties of
each. In this subsection, we discuss the global implications of
several facets of the dark-matter halo response, focusing first on
changes to the density and velocity structure, effects that might be
traced by halo stars (e.g. observable by Gaia), and then turn to
implications for the fast bar--slow bar formation scenarios.

\begin{figure}
  \centering 
  \includegraphics[width=3.5in]{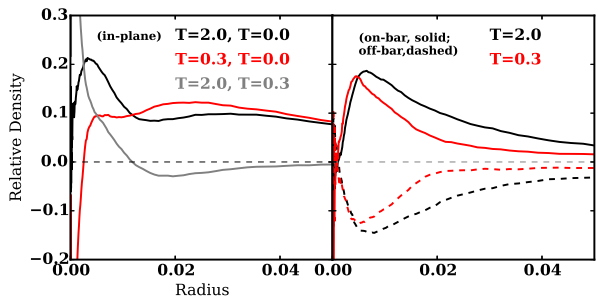} 
  \caption{Left panel: the in-plane dark matter density ratio for axisymmetric
    density profiles at three pairs of times.  The black line compares
    $T=2.0$ with $T=0.0$ (the total change in the profile during the
    simulation), the red line compares $T=0.3$ to $T=0.0$
    (the change in the halo distribution induced by bar formation), and
    the grey line compares $T=2.0$ to $T=0.3$ (changes caused by
    secular evolution). The halo has dramatically redistributed its
    density in response to the presence of a stellar disc by $T=0.3$,
    but continues to evolve secularly owing to the presence of the
    bar. Outside of $R=0.05$, the fractional density difference
    rapidly falls to zero. Right panel: the in-plane fractional dark matter
    density
    ratio parallel to the bar and perpendicular to the bar,
    compared to the axisymmetrised profile, for two times. The on-bar
    radial cut shows an enhancement at a $>$5 per cent level within
    $R=0.02$ following bar formation ($T=0.3$), and a $>10$ per cent
    enhancement within $R=0.02$ after further evolution
    ($T=2.0$). Radius and time are given in virial units, as in Fig.~\ref{fig:rotcurve_evolved}.
    \label{fig:halodens}}
\end{figure}

\subsubsection{Density and velocity structure} \label{subsubsec:denvelstruct}

The bar changes the density and velocity structure of the dark matter
halo. The dark disc in the fiducial simulation
appears as an in-plane overdensity that grows with time. However, the
non-axisymmetric structure of the shadow bar also contributes to
density variations in the plane. Figure~\ref{fig:halodens} demonstrates
these two effects at play. In the left panel, we plot the
axisymmetric density (azimuthally averaged) ratios for three pairs of
times to directly explore the temporal evolution of the planar density
without the shadow bar. Over the course of the entire simulation, from
$T=0$ to $T=2$ ($\delta T=4$ Gyr scaled to the MW; black curve), the
in-plane density out to $R=0.05$ (15 kpc for the MW) increases everywhere
by $>8$ per cent . Inside of $R=0.01$, the density increases by up to
20 per cent. For comparison, the evolution from $T=0$ to $T=0.3$ is
shown in red, demonstrating that the majority of the azimuthally
averaged evolution at $R>0.015$ (4.5~kpc for the MW) happens as the bar
is forming. The evolution from $T=0.3$ to $T=2$ ($\delta T=3.4$~Gyr
for the MW), shown in grey, corroborates this, and shows that the in-plane
density at $R<0.015$ increases dramatically as the bar continues to
grow.

The right-hand panel of Figure~\ref{fig:halodens} examines the density
on and off the bar axis, relative to the axisymmetric density, at two
different times. The dark matter distribution is significantly
non-axisymmetric at $T=0.3$, in the midst of bar formation, including
an enhancement of up to 17 per cent at $R=0.005$ (1.5~kpc for the MW). At
$T=2.0$, the asymmetry is more pronounced enhancing the 
overdensity in the direction of the bar 
to 10 per cent out to $R=0.02$ (6~kpc for the MW) and to 5 per cent
out to $R=0.035$ (10.5~kpc for the MW),
as well as shifting the peak overdensity to a larger radius as
compared to $T=0.3$.
Although the azimuthally
averaged profile remains roughly unchanged at $R>0.015$, the
bar produces significant non-axisymmetric halo evolution at all relevant
disc radii.

\begin{figure}
  \centering 
  \includegraphics[width=3.5in]{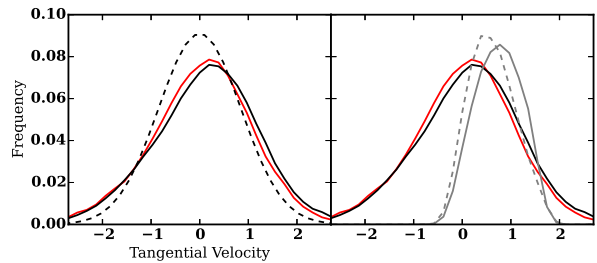} 
  \caption{Left panel: Tangential velocity distributions at $T=2.0$ on- 
    (solid black)
    and off-bar (red) for the
    dark matter halo, at $R=0.008$ (near the end of the bar). The
    dashed black line indicates a normal distribution.
    The tangential velocity distribution on the
    bar axis (solid black) shows both an enhancement centred near $V_c=0.8 (120
    ~{\rm km~s^{-1}}$ for the MW $)$, approximately
    the bar velocity at this radius, as well as a shift in the overall
    velocity peak. Both curves show enhanced tails relative to the
    normal distribution, at
    velocities $V_c<-1.5$. Right panel: The same dark halo tangential velocity
    distributions as the left panel are plotted, along with the disc on-bar
    distribution and axisymmetric distribution shown as the solid grey
    and dashed grey lines, respectively (scaled by 0.6). Tangential
    velocity is given in virial units, as in Fig.~\ref{fig:rotcurve}.\label{fig:halovels}}
\end{figure}

Further, the isotropic nature of the dark matter halo velocity in the
plane of the disc has been significantly altered through an infusion
of $L_{\rm z}$ from the disc. In the left panel of
Figure~\ref{fig:halovels}, we plot the tangential velocity distributions
on- and off-bar for particles (black and red curves, respectively)
near the end of the bar ($0.008 < R <0.01$) at $T=2$. For comparison, the
dashed black line indicates a normal distribution with the same
tangential velocity dispersion, centred at zero velocity. Along the
bar axis, the peak of the tangential velocity ($V_\theta$)
distribution is shifted by $\delta V_\theta= 0.2$ (30$~{\rm
  km~s^{-1}}$ for the MW). Even off the bar axis, the peak of the
tangential velocity distribution has been shifted by $\delta V_\theta=
0.1~$ (15$~{\rm km~s^{-1}}$ for the MW). Additionally, the tails of both the
on- and off-bar distributions are enhanced relative to a normal
distribution owing to secular processes reshaping the halo.

In the right panel of Figure~\ref{fig:halovels} we again plot the
tangential velocity distribution for the on- and off-bar populations, but
now compare it to the disc. We plot the on-bar disc distribution in
the same annulus near
the end of the bar as the solid grey line and the axisymmetric velocity
distribution of velocities in the disc as the dashed grey
line\footnote{5 per cent of the disc orbits at the end of the bar are
  retrograde, reflecting precession toward the bar that causes the
  tangential velocity in the inertial frame to be negative.}. The
on-bar disc particles are shifted to higher tangential
velocities ($V_{\theta, {\rm bar}}=0.8 = 120 ~{\rm km~s^{-1}}$ for the
MW, including a shift between peaks of $\delta V_\theta=0.5= 75~{\rm km~s^{-1}}$ for the
MW), which the halo reflects, albeit at a smaller $\delta V$ (as
discussed above). The on-bar disc distribution also shows an
enhancement relative to the axisymmetric distribution at $V_\theta=1.4$ that is mildly reflected in
the halo. The particles trapped into the bar potential librate
around the pattern speed as the flattening and trapping of dark matter particles correlate
the velocities to create a non-zero mean velocity distribution that is
particularly enhanced along the bar axis (in addition to the generic
rotation seen in the off-bar population). In
other words, the the $L_z$ distribution has been modified, with the
enhancement most clearly seen along the bar major axis. We make predictions for the net effect of the above considerations on
dark matter direct detection experiments in \citet{petersen15a}.

Although our fiducial model is motivated by $\Lambda$CDM simulations
(halo) and direct observations (disc), it represents a single
realisation from a particular model family.  We emphasise that these
observational conclusions are likely to vary even within a model
family.  However, we expect that the dynamical nature of the density and
velocity differences driven by the shadow bar will be generic. To reduce the uncertainty of these simulation-derived quantities for
application to Earth-based dark matter detection experiments, at least
three additional pieces of information are critical.  First, both the
density and velocity signatures will be affected by the assumed length
of the MW bar and its evolution as a function of time.  Relative to
the disc scale length, our bar is 
shorter than the observed MW bar \citep{wegg15}; 
also, the MW bar may have been longer in the past
depending on its formation mechanism. Therefore, dark matter at the solar
radius maybe have been influenced more readily. Secondly, we have
demonstrated that the velocity structure depends on the position
angle of the bar, so we need an accurate determination of the MW bar
position angle. Finally, we need the formation time of the MW bar. The shadow bar continues to grow in time relative to the mass
of the stellar bar.  Therefore, an older MW bar would have more
trapped dark matter material relative to the stellar bar. An older MW bar may also have a different velocity
structure due to trapped dark matter orbits
contributing to the torque on the untrapped halo component and the efficiency of $L_z$
transfer from the enhanced mass of the shadow bar.

\subsubsection{Angular momentum transport} 
\label{subsubsec:angmomtrans}

\begin{figure} 
  \centering 
  \includegraphics[width=3.5 in]{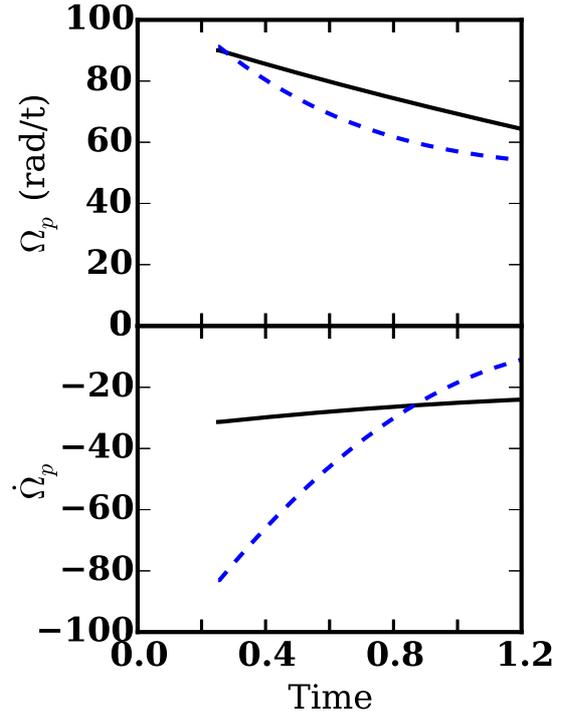} 
  \caption{Upper panel: the pattern speed for the fiducial simulation
    (black) and an azimuthally-shuffled halo to suppress shadow bar
    formation (Fs; the dashed blue line). The
    simulation time is limited to $T<1.2$ so that the stellar bars
    that form are still roughly equivalent. Lower panel: Change in
    pattern speed, following the same colour scheme. The shuffled
    simulation slows much more rapidly during the simulation,
    indicative of an increased $L_{\rm z}$ acceptance by the
    halo when there is no shadow bar. Time is given in virial units, as in Fig.~\ref{fig:patternspeed}.\label{fig:shufquan}}
\end{figure}

The shadow bar contains a large fraction of the dark-matter orbits at
low inclination: $>45$ per cent  of dark matter orbits satisfying $0.003<R<0.01$ and $0.9<\cos\beta\le
1$ are trapped, as discussed in section~\ref{subsubsec:haloorbit}. The
low inclination orbits are the same orbits
responsible for accepting angular momentum and slowing the bar as
determined by examining the model of \citet{weinberg85}.  We expected
that this would diminish the torque relative to the canonical
perturbation theory results (e.g. \citealt{tremaine84}).  We test this
speculation by azimuthally shuffling the halo orbits as described in
section \ref{subsubsec:numericalvariants}.  The purpose of this
simulation is to prevent the shadow bar from forming and mimic the
standard dark-matter halo dynamical friction scenario.
Figure~\ref{fig:shufquan} compares the results of the fiducial
simulation to the azimuthally shuffled experiment.  The
azimuthally-shuffled halo is able to transfer angular momentum
continuously, evident through the rapidly decreasing pattern speed
during the evolution.  Given the assumption that $\Delta
L\approx\Delta \Omega_p$ for a stellar bar with the same geometry, we
find that $\Delta\Omega_{p, {\rm shuffled}}/\Delta\Omega_{p, {\rm
    self-consistent}} \approx 3$, an appreciable increase in torque. Clearly, inhibiting the formation of the shadow
bar results in a strong torque on the bar by the halo.
We only present the results of the simulation for
$T<1.2$ because after this time the resultant stellar bars no longer
qualitatively resemble each other ($I_{\rm bar, shuffled}\ne I_{\rm bar,
  self-consistent}$ where $I$ is the moment of inertia of the bar), making the comparison of the
simulations unfair.  

The shadow bar does not eliminate the torque altogether, of course.  A
massive bar, enhanced by the dark matter, will couple more strongly to
higher order resonances, even though these resonances are much
weaker. The relative strength can be roughly approximated by
considering the inverse of the largest winding number in the resonant
equation (the maximum of $[m,~l_\phi,~l_r]$ in equation \ref{eq:resonances}) for
a particular resonance. For example, the 4:2:1 resonance will be
approximately half as strong as the 2:2:1 resonance, and so on, for all
realistic distribution functions \citep{binney08}. However, the
primary (low-order) resonances are weakened by the reduction in
available dark matter to accept angular momentum as more halo material
becomes trapped. An enhanced dark disc, fed through satellite
  accretion, could act as a larger reservoir for material to trap,
  with presently unexplored implications for the torque.

\section{Comparison between our fiducial and other models}\label{sec:discussion}

\begin{figure*} 
  \centering 
  \includegraphics[width=6 in]{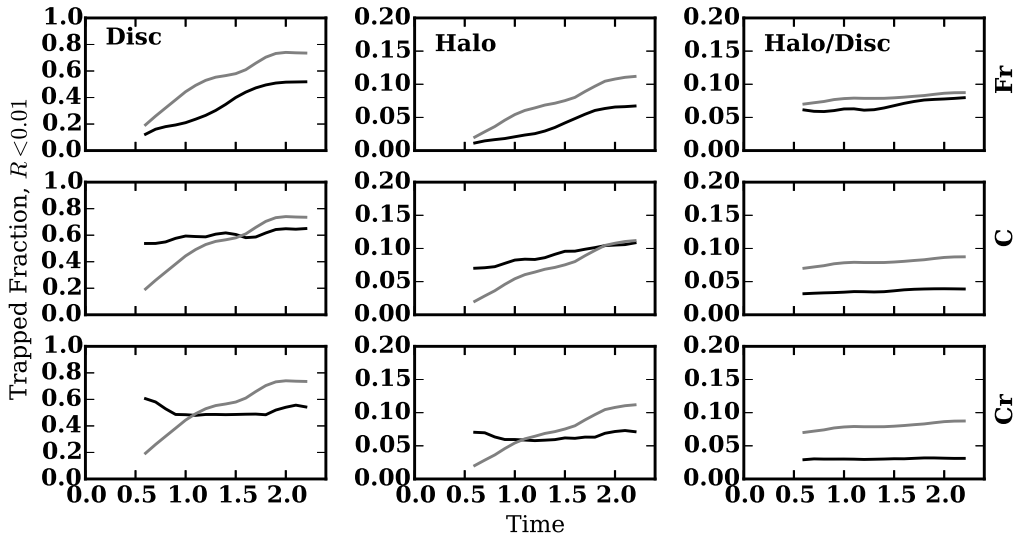} 
  \caption{From left to right: stellar disc trapping, dark matter
    trapping, and the ratio of dark-matter halo to stellar disc trapping
    by the bar inside $R=0.01$. Compare these to
    Fig.~\ref{fig:disctrap}. From top to bottom the simulations are:
    the cusped rotating halo (Fr in Table \ref{tab:simlist}), the
    cored nonrotating (C), and the cored rotating (Cr) models. 
    Black lines indicate the total mass trapped scaled by the
    mass interior to $R=0.01$ and grey lines indicate the corresponding curve for the fiducial
    model. Time is given in virial units, as in Fig.~\ref{fig:disctrap}.\label{fig:timetrap}}

\end{figure*}

\subsection{Halo initial conditions} \label{subsec:haloinitials} 

Our fiducial model is a simple representation of a disc galaxy in the
$\Lambda$CDM formation scenario: a cuspy dark-matter halo profile with
an exponential stellar disc.  However, there are a variety of
alternatives that are commonly explored in the literature. These
include models whose density approaches a constant at some
characteristic core radius $r_c$ and rotating halo models (models with
a non-zero value of $L_z$).  The former, which have some observational
(e.g. rotation curves, \citealt{breddels13}) and theoretical
motivation (e.g. satellite accretion producing cores, black holes, and
feedback from star formation or supernova; \citealt{tremaine99},
\citealt{mashchenko06}, \citealt{governato12}) have often been 
used in N-body bar simulations (\citealt{athanassoula02},
\citealt{martinez06}, \citealt{saha10}).  The latter are motivated
by both evolutionary processes such as merging \citep{dekel03} and the
hierarchical formation process itself (\citealt{navarro97}, \citealt{bullock01}).  In this section, we
briefly summarise the variation in the properties of the disc and
shadow bars for a representative member of each of these two model
classes.

Table \ref{tab:simlist} describes the halo models we tested in
addition to the fiducial cuspy non-rotating model. The resultant disc
and shadow bar evolutions are shown in Figure~\ref{fig:timetrap}, the
analogue of Figure~\ref{fig:disctrap}. The qualitative evolution is
similar between all four simulations, suggesting that the results
presented in section \ref{sec:results} are generally relevant.  In
particular, the shadow bar continues to grow relative to the stellar
bar in all the simulations (the right panel of
Figure~\ref{fig:timetrap}). The fiducial model exhibits the strongest
stellar and shadow bars at late times as a fraction of the available mass, but the trend in each model suggests that an
equilibrium has not been reached at $T=2.0$, the end of the
simulations.  (Recall that one system time unit is approximately 2
Gyr scaled to the MW, so that $T=2.0$ is roughly 4 Gyr or 40 per cent of
the age of the Galaxy.)  The cored halo models exhibit a buckling
instability at $T=0.6$ in the cored nonrotating model, and at $T=1.0$ in
the cored rotating model. This instability releases disc orbits trapped
during the bar formation phase, albeit a relatively small fraction,
and does not appear to reduce the dark-matter fraction relative to the
trapped disc fraction.

Overall, the differences in evolution between the models can be
described as follows: (1) the cored halo increases the initial
strength of the stellar bar through a stronger instability (rapid
growth at $T<0.6$).  However, the relative strength of the shadow bar
is smaller and stagnant at late times owing to the reduced central
density and, consequently, a smaller population of halo orbits that are
available to trap; and (2) the increased rotation of a halo affects
models by reducing the strength of the bar at early times (diminished the
growth between $T=0.5$ and $T=1.0$); the stellar and
shadow bars take longer to dynamically mature and, consequently, have not
evolved to be as massive over the same span of time. This effect is
smaller in the cored rotating halo compared to the cusped rotating
halo, suggesting that the cuspy or cored nature of the halo dominates
the evolution when compared to rotation in these models. We will
discuss these trends further in a future paper.

\subsection{Disc Maximality} \label{subsec:discmax}

As discussed in section~\ref{subsubsec:fiducial}, our fiducial model
is a submaximal disc (as are all F-series models), chosen to correspond to
cosmological simulations. Simulations with maximal discs (such as the C-series models) are subject
to a violent disc instability that results in a qualitatively
different bar formation scenario than that of our fiducial F-series
models\footnote{The C-series models have a significantly reduced
  fraction of dark matter within a scalelength, initially $M_{\rm
    dm}/M_{\star}(R<0.01)=0.332$, reducing to $M_{\rm
    dm}/M_{\star}(R<0.01)=0.231$ at $T=2.0$ (owing to an increase in
  disc mass at $R<0.01$). Compare this to values for the fiducial model in Section~\ref{subsec:orbitstat}.}. The
violent disc instability may create a noisy bar formation process that
inhibits the formation of the dark disc required for the stellar bar
to then trap into the shadow bar. The C-series models nearly reach
their maximum trapping fraction 
within several dynamical times of formation where the F-series models
trap continuously over many dynamical times (see
Figure~\ref{fig:timetrap}). This illustrates a the distinct
differences between the fast and slow mode of bar growth in our models. The variations with time, including
untrapping of some orbits, suggest that
the trapping is more tenuous in a violent instability. The fast and
slow modes of bar growth will be studied further in future work. If these differences can be detected in an out-of-plane luminous tracer population (e.g. spheroid or thick disc stars in the Milky Way), we may be able to discriminate between these two modes of formation in nature.

It is also likely that the relative contributions to the potential from the
disc and halo will change the formation efficiency of the shadow bar,
an effect we have already seen qualitatively in the contrast between
the cusp and core simulations. In the simulations with cores, the halo
material interior to $R_d$ is approximately half that in the cusp
simulation, so that even when the same fraction of material is
trapped (e.g. the middle panel of the cored simulation in
Figure~\ref{fig:timetrap}), the contribution to the overall bar
potential is low (e.g. the right panel of the cored simulation in
Figure~\ref{fig:timetrap}). Furthermore, a more violent formation of the bar may change the
structure of the dark matter wake. In this bar formation scenario, the formation of the dark disc and subsequent shadow
bar trapping, both of which are slow processes, could be disrupted.  A study of
the implications of stellar disc maximality for the shadow bar and
dark matter wake will be investigated in future work.

\section{Conclusion} 
\label{sec:conclusion} 

Our main findings are as follows:
\begin{enumerate}
\renewcommand{\theenumi}{(\arabic{enumi})}
\item The stellar bar traps dark matter into a \emph{shadow bar}. This
  shadow bar has a mass that is $>$6 per cent of the stellar bar mass
  after the bar forms, and this ratio increases with time throughout
  the simulations to values $>$9 per cent by mass.  The halo is deformed by
  the presence of the disc as well as the continued torque from the stellar
  and shadow bar, creating a population of disc-like orbits (with a
  preferred angular momentum axis).  We suggest that
  this reservoir increases the trapping rate of dark matter orbits.
\item Trapping in the dark matter halo and the stellar disc takes
  place both at bar formation and throughout the simulation.
\item The existence and strength of the shadow bar does not change
  appreciably in the presence of a core or rotation in the
  halo. However, the trapping rate of both the dark matter halo and
  the stellar component strongly depends on halo profile and the initial
  angular momentum distribution in the halo.
\item The dark matter halo exhibits a density and velocity signature
  indicative of a reaction to the presence of the stellar disc at
  radii much larger than that of the bar radius. The density and
  velocity structure change with the angle to the bar, even at
  several bar radii. This could have important implications for direct
  dark matter detection experiments.
\item Approximately 12 per cent of the total halo inside of the bar
  radius ($R<0.01$), and 45 per cent of the dark matter satisfying
  $\cos\beta>0.9$ within $0.003<R<0.01$, is trapped after
  4 Gyr. These trapped orbits are precisely the ones that dominate the
  secular angular momentum transfer to the halo in the absence of
  trapping. The trapping changes the secular angular momentum
  transfer rates that one would estimate without trapping. We
  demonstrate this point with a simulation that suppresses trapped
  orbits by artificially decorrelating the halo orbit apsides with the
  bar position.  This increases the halo torque on the bar after
  formation by a factor of three!  Stronger bars are likely
  proportionally larger fractions of their haloes.  This suggests that
  halo torques may be much smaller than theoretically predicted,
  especially for strong bars, helping to resolve the tension between
  the $\Lambda$CDM scenario and the observational evidence for rapidly
  rotation bars.
\end{enumerate}

In concept, the idea of trapped dark matter is striking in its
simplicity: when disc orbits and halo orbits occupy similar places
in phase space, they will respond similarly to a perturbation.  We have
only begun to explore the parameter space necessary to characterise
dark matter trapping, as well as the implications for the evolution of
isolated galaxies.  Therefore, we expect that other simulations have
shadow bars as well, but previous studies have not investigated or
identified the halo trapping and have not undertaken a detailed study
of trapped populations in the disc component. Our $k$-means orbit
finding algorithm
can be used as a technique to further understand the dynamics of
strongly time-dependent evolution, such as the bar-disc-halo
system. In future papers, we will present a detailed breakdown of the
orbit populations in the disc and halo components using the $k$-means
technique.

Many simulations have shown that galactic bars efficiently transport
angular momentum from the disc to the halo, causing the bar to slow,
lengthen, and increase in strength (\citealt{debattista00};
\citealt{athanassoula02}; \citealt{debattista06};
\citealt{weinberg07b}).  Many of these same authors have argued that
bars are long-lived, as no mechanism for their destruction has been
observed in simulations of isolated systems
(e.g. \citealt{athanassoula13}). The results of \citet{hernquist92}
then suggest that the bar either cannot be strongly coupled to the
halo, or requires fine-tuning to eliminate lower-order resonances to
reduce the strong torque. The suggestion that bars cannot be strongly
coupled to the halo seems to be in direct conflict with the angular momentum
transfer rates observed in simulations depicting a rapidly slowing
bar.  The production of a shadow bar, and the trapping of halo orbits
by the bar more generally, offers a possible way out of this 
dynamical conundrum.  Without a sink of halo particles that are
distributed asymmetrically in phase space relative to the bar, there
can be no torque.  Our discussion in section \ref{subsec:haloinitials}
suggests that the existence of the shadow bar is robust, but that its
magnitude will depend on the galaxy profile and the details of the
subsequent bar formation. If a significant dark disc exists as the
  remnant of shredded satellites accreted by the galaxy, the
  magnitude of the shadow bar could be significantly enhanced.

Although the shadow bar is kinematically similar to the stellar bar,
the findings presented here provide new insight into the kinematics of
the galactic centre. Observationally, stellar surveys may detect
rotation in the stellar halo toward the inner Galaxy that would be
indicative of trapping in a spherical component. For stars and stellar
systems that may have formed prior to the formation of the MW bar
(such as metal-poor halo stars, thick-disc stars, or globular
clusters), the findings presented here indicate that these components
could also be trapped into the bar potential. Suggestions in the literature
dating as early as \citet{blitz91} of a triaxial rotating component at
the centre of the galaxy could be the result of an originally spheroidal
component trapped by the bar potential. Recent observations by
\citet{babusiaux14} and \citet{rossi15} have pointed out kinematic
signatures (e.g. an anomalous tangential and radial velocity
dispersion) that could be consistent with this interpretation. New
stellar survey data available in the coming years will provide further
insights. The results of this paper suggest that stellar survey data
may be able to discriminate between fast and slow-growth bar
formation and evolution in the MW.

Finally, evidence for a shadow bar in our own Milky Way could be
directly detected through future direct dark matter detection
experiments, including directional direct detection experiments, which
would provide significant dynamical insights into the unseen dark
matter component.

\section*{Acknowledgments}

We thank the anonymous referee for
helpful suggestions that strengthened this paper. We also thank Victor Debattista and Jerry Sellwood for constructive
comments on the preprint that clarified the main points. 

\bibliography{PetersenM}

\begin{thebibliography}{}
\makeatletter
\relax
\def\mn@urlcharsother{\let\do\@makeother \do\$\do\&\do\#\do\^\do\_\do\%\do\~}
\def\mn@doi{\begingroup\mn@urlcharsother \@ifnextchar [ {\mn@doi@}
  {\mn@doi@[]}}
\def\mn@doi@[#1]#2{\def\@tempa{#1}\ifx\@tempa\@empty \href
  {http://dx.doi.org/#2} {doi:#2}\else \href {http://dx.doi.org/#2} {#1}\fi
  \endgroup}
\def\mn@eprint#1#2{\mn@eprint@#1:#2::\@nil}
\def\mn@eprint@arXiv#1{\href {http://arxiv.org/abs/#1} {{\tt arXiv:#1}}}
\def\mn@eprint@dblp#1{\href {http://dblp.uni-trier.de/rec/bibtex/#1.xml}
  {dblp:#1}}
\def\mn@eprint@#1:#2:#3:#4\@nil{\def\@tempa {#1}\def\@tempb {#2}\def\@tempc
  {#3}\ifx \@tempc \@empty \let \@tempc \@tempb \let \@tempb \@tempa \fi \ifx
  \@tempb \@empty \def\@tempb {arXiv}\fi \@ifundefined
  {mn@eprint@\@tempb}{\@tempb:\@tempc}{\expandafter \expandafter \csname
  mn@eprint@\@tempb\endcsname \expandafter{\@tempc}}}

\bibitem[\protect\citeauthoryear{{Aalseth} et~al.,}{{Aalseth}
  et~al.}{2013}]{aalseth13}
{Aalseth} C.~E.,  et~al., 2013, \mn@doi [\prd] {10.1103/PhysRevD.88.012002},
  \href {http://adsabs.harvard.edu/abs/2013PhRvD..88a2002A} {88, 012002}

\bibitem[\protect\citeauthoryear{{Agnese} et~al.,}{{Agnese}
  et~al.}{2013}]{agnese13}
{Agnese} R.,  et~al., 2013, \mn@doi [\prd] {10.1103/PhysRevD.88.031104}, \href
  {http://adsabs.harvard.edu/abs/2013PhRvD..88c1104A} {88, 031104}

\bibitem[\protect\citeauthoryear{{Aguerri} et~al.,}{{Aguerri}
  et~al.}{2015}]{aguerri15}
{Aguerri} J.~A.~L.,  et~al., 2015, \mn@doi [\aap]
  {10.1051/0004-6361/201423383}, \href
  {http://adsabs.harvard.edu/abs/2015A%26A...576A.102A} {576, A102}

\bibitem[\protect\citeauthoryear{{Allende Prieto} et~al.,}{{Allende Prieto}
  et~al.}{2008}]{allende08}
{Allende Prieto} C.,  et~al., 2008, \mn@doi [Astronomische Nachrichten]
  {10.1002/asna.200811080}, \href
  {http://adsabs.harvard.edu/abs/2008AN....329.1018A} {329, 1018}

\bibitem[\protect\citeauthoryear{{Allgood}, {Flores}, {Primack}, {Kravtsov},
  {Wechsler}, {Faltenbacher}  \& {Bullock}}{{Allgood} et~al.}{2006}]{allgood06}
{Allgood} B.,  {Flores} R.~A.,  {Primack} J.~R.,  {Kravtsov} A.~V.,  {Wechsler}
  R.~H.,  {Faltenbacher} A.,   {Bullock} J.~S.,  2006, \mn@doi [\mnras]
  {10.1111/j.1365-2966.2006.10094.x}, \href
  {http://adsabs.harvard.edu/abs/2006MNRAS.367.1781A} {367, 1781}

\bibitem[\protect\citeauthoryear{{Athanassoula}}{{Athanassoula}}{1992}]{athanassoula92}
{Athanassoula} E.,  1992, \mnras, \href
  {http://adsabs.harvard.edu/abs/1992MNRAS.259..328A} {259, 328}

\bibitem[\protect\citeauthoryear{{Athanassoula}}{{Athanassoula}}{2007}]{athanassoula07}
{Athanassoula} E.,  2007, \mn@doi [\mnras] {10.1111/j.1365-2966.2007.11711.x},
  \href {http://adsabs.harvard.edu/abs/2007MNRAS.377.1569A} {377, 1569}

\bibitem[\protect\citeauthoryear{Athanassoula \& Misiriotis}{Athanassoula \&
  Misiriotis}{2002}]{athanassoula02}
Athanassoula E.,  Misiriotis A.,  2002, MNRAS, 330, 35

\bibitem[\protect\citeauthoryear{{Athanassoula}, {Machado}  \&
  {Rodionov}}{{Athanassoula} et~al.}{2013}]{athanassoula13}
{Athanassoula} E.,  {Machado} R.~E.~G.,   {Rodionov} S.~A.,  2013, \mn@doi
  [\mnras] {10.1093/mnras/sts452}, \href
  {http://adsabs.harvard.edu/abs/2013MNRAS.429.1949A} {429, 1949}

\bibitem[\protect\citeauthoryear{{Babusiaux} et~al.,}{{Babusiaux}
  et~al.}{2014}]{babusiaux14}
{Babusiaux} C.,  et~al., 2014, \mn@doi [\aap] {10.1051/0004-6361/201323044},
  \href {http://adsabs.harvard.edu/abs/2014A%26A...563A..15B} {563, A15}

\bibitem[\protect\citeauthoryear{{Barnes}, {Hut}  \& {Goodman}}{{Barnes}
  et~al.}{1986}]{barnes86}
{Barnes} J.,  {Hut} P.,   {Goodman} J.,  1986, \mn@doi [\apj] {10.1086/163786},
  \href {http://adsabs.harvard.edu/abs/1986ApJ...300..112B} {300, 112}

\bibitem[\protect\citeauthoryear{{Bernabei} et~al.,}{{Bernabei}
  et~al.}{2014}]{bernabei14}
{Bernabei} R.,  et~al., 2014, \mn@doi [Nuclear Instruments and Methods in
  Physics Research A] {10.1016/j.nima.2013.10.079}, \href
  {http://adsabs.harvard.edu/abs/2014NIMPA.742..177B} {742, 177}

\bibitem[\protect\citeauthoryear{Binney \& Spergel}{Binney \&
  Spergel}{1982}]{binney82}
Binney J.,  Spergel D.,  1982, ApJ, 252, 308

\bibitem[\protect\citeauthoryear{{Binney} \& {Tremaine}}{{Binney} \&
  {Tremaine}}{2008}]{binney08}
{Binney} J.,  {Tremaine} S.,  2008, {Galactic Dynamics: Second Edition}.
Princeton University Press

\bibitem[\protect\citeauthoryear{{Blitz} \& {Spergel}}{{Blitz} \&
  {Spergel}}{1991}]{blitz91}
{Blitz} L.,  {Spergel} D.~N.,  1991, \mn@doi [\apj] {10.1086/170535}, \href
  {http://adsabs.harvard.edu/abs/1991ApJ...379..631B} {379, 631}

\bibitem[\protect\citeauthoryear{{Bournaud} \& {Combes}}{{Bournaud} \&
  {Combes}}{2002}]{bournaud02}
{Bournaud} F.,  {Combes} F.,  2002, \mn@doi [\aap]
  {10.1051/0004-6361:20020920}, \href
  {http://adsabs.harvard.edu/abs/2002A%26A...392...83B} {392, 83}

\bibitem[\protect\citeauthoryear{{Bournaud}, {Combes}  \& {Semelin}}{{Bournaud}
  et~al.}{2005}]{bournaud05}
{Bournaud} F.,  {Combes} F.,   {Semelin} B.,  2005, \mn@doi [\mnras]
  {10.1111/j.1745-3933.2005.00096.x}, \href
  {http://adsabs.harvard.edu/abs/2005MNRAS.364L..18B} {364, L18}

\bibitem[\protect\citeauthoryear{{Bovy} \& {Rix}}{{Bovy} \&
  {Rix}}{2013}]{bovy13}
{Bovy} J.,  {Rix} H.-W.,  2013, \mn@doi [\apj] {10.1088/0004-637X/779/2/115},
  \href {http://adsabs.harvard.edu/abs/2013ApJ...779..115B} {779, 115}

\bibitem[\protect\citeauthoryear{{Boylan-Kolchin}, {Bullock}, {Sohn}, {Besla}
  \& {van der Marel}}{{Boylan-Kolchin} et~al.}{2013}]{boylan13}
{Boylan-Kolchin} M.,  {Bullock} J.~S.,  {Sohn} S.~T.,  {Besla} G.,   {van der
  Marel} R.~P.,  2013, \mn@doi [\apj] {10.1088/0004-637X/768/2/140}, \href
  {http://adsabs.harvard.edu/abs/2013ApJ...768..140B} {768, 140}

\bibitem[\protect\citeauthoryear{{Breddels} \& {Helmi}}{{Breddels} \&
  {Helmi}}{2013}]{breddels13}
{Breddels} M.~A.,  {Helmi} A.,  2013, \mn@doi [\aap]
  {10.1051/0004-6361/201321606}, \href
  {http://adsabs.harvard.edu/abs/2013A%26A...558A..35B} {558, A35}

\bibitem[\protect\citeauthoryear{{Bruch}, {Read}, {Baudis}  \& {Lake}}{{Bruch}
  et~al.}{2009}]{bruch09}
{Bruch} T.,  {Read} J.,  {Baudis} L.,   {Lake} G.,  2009, \mn@doi [\apj]
  {10.1088/0004-637X/696/1/920}, \href
  {http://adsabs.harvard.edu/abs/2009ApJ...696..920B} {696, 920}

\bibitem[\protect\citeauthoryear{Bullock, Dekel, Kolatt, Kravtsov, Klypin,
  Porcianai  \& Primack}{Bullock et~al.}{2001}]{bullock01}
Bullock J.,  Dekel A.,  Kolatt T.,  Kravtsov A.,  Klypin A.,  Porcianai C.,
  Primack J.,  2001, ApJ, 555, 240

\bibitem[\protect\citeauthoryear{{Cheung} et~al.,}{{Cheung}
  et~al.}{2013}]{cheung13}
{Cheung} E.,  et~al., 2013, \mn@doi [\apj] {10.1088/0004-637X/779/2/162}, \href
  {http://adsabs.harvard.edu/abs/2013ApJ...779..162C} {779, 162}

\bibitem[\protect\citeauthoryear{{Clutton-Brock}}{{Clutton-Brock}}{1973}]{cluttonbrock73}
{Clutton-Brock} M.,  1973, \mn@doi [\apss] {10.1007/BF00647652}, \href
  {http://adsabs.harvard.edu/abs/1973Ap%26SS..23...55C} {23, 55}

\bibitem[\protect\citeauthoryear{{Clutton-Brock}, {Innanen}  \&
  {Papp}}{{Clutton-Brock} et~al.}{1977}]{cluttonbrock77}
{Clutton-Brock} M.,  {Innanen} K.~A.,   {Papp} K.~A.,  1977, \mn@doi [\apss]
  {10.1007/BF00642839}, \href
  {http://adsabs.harvard.edu/abs/1977Ap%26SS..47..299C} {47, 299}

\bibitem[\protect\citeauthoryear{{Col{\'{\i}}n}, {Valenzuela}  \&
  {Klypin}}{{Col{\'{\i}}n} et~al.}{2006}]{colin06}
{Col{\'{\i}}n} P.,  {Valenzuela} O.,   {Klypin} A.,  2006, \mn@doi [\apj]
  {10.1086/503791}, \href {http://adsabs.harvard.edu/abs/2006ApJ...644..687C}
  {644, 687}

\bibitem[\protect\citeauthoryear{Contopoulos \& Papayannopoulos}{Contopoulos \&
  Papayannopoulos}{1980}]{contopoulos80}
Contopoulos G.,  Papayannopoulos T.,  1980, A\&A, 92, 33

\bibitem[\protect\citeauthoryear{{Debattista} \& {Sellwood}}{{Debattista} \&
  {Sellwood}}{2000}]{debattista00}
{Debattista} V.~P.,  {Sellwood} J.~A.,  2000, \mn@doi [\apj] {10.1086/317148},
  \href {http://adsabs.harvard.edu/abs/2000ApJ...543..704D} {543, 704}

\bibitem[\protect\citeauthoryear{Debattista, Mayer, Carollo, Moore, Wadsley  \&
  Quinn}{Debattista et~al.}{2006}]{debattista06}
Debattista V.,  Mayer L.,  Carollo C.,  Moore B.,  Wadsley J.,   Quinn T.,
  2006, ApJ, 645, 209

\bibitem[\protect\citeauthoryear{{Debattista}, {Moore}, {Quinn}, {Kazantzidis},
  {Maas}, {Mayer}, {Read}  \& {Stadel}}{{Debattista}
  et~al.}{2008}]{debattista08}
{Debattista} V.~P.,  {Moore} B.,  {Quinn} T.,  {Kazantzidis} S.,  {Maas} R.,
  {Mayer} L.,  {Read} J.,   {Stadel} J.,  2008, \mn@doi [\apj]
  {10.1086/587977}, \href {http://adsabs.harvard.edu/abs/2008ApJ...681.1076D}
  {681, 1076}

\bibitem[\protect\citeauthoryear{{Dekel}, {Devor}  \& {Hetzroni}}{{Dekel}
  et~al.}{2003}]{dekel03}
{Dekel} A.,  {Devor} J.,   {Hetzroni} G.,  2003, \mn@doi [\mnras]
  {10.1046/j.1365-8711.2003.06432.x}, \href
  {http://adsabs.harvard.edu/abs/2003MNRAS.341..326D} {341, 326}

\bibitem[\protect\citeauthoryear{{Earn}}{{Earn}}{1996}]{earn96}
{Earn} D.~J.~D.,  1996, \mn@doi [\apj] {10.1086/177404}, \href
  {http://adsabs.harvard.edu/abs/1996ApJ...465...91E} {465, 91}

\bibitem[\protect\citeauthoryear{{Gilmore} et~al.,}{{Gilmore}
  et~al.}{2012}]{gilmore12}
{Gilmore} G.,  et~al., 2012, The Messenger, \href
  {http://adsabs.harvard.edu/abs/2012Msngr.147...25G} {147, 25}

\bibitem[\protect\citeauthoryear{{Governato} et~al.,}{{Governato}
  et~al.}{2012}]{governato12}
{Governato} F.,  et~al., 2012, \mn@doi [\mnras]
  {10.1111/j.1365-2966.2012.20696.x}, \href
  {http://adsabs.harvard.edu/abs/2012MNRAS.422.1231G} {422, 1231}

\bibitem[\protect\citeauthoryear{{Hernquist} \& {Ostriker}}{{Hernquist} \&
  {Ostriker}}{1992}]{hernquist92a}
{Hernquist} L.,  {Ostriker} J.~P.,  1992, \mn@doi [\apj] {10.1086/171025},
  \href {http://adsabs.harvard.edu/abs/1992ApJ...386..375H} {386, 375}

\bibitem[\protect\citeauthoryear{{Hernquist} \& {Weinberg}}{{Hernquist} \&
  {Weinberg}}{1992}]{hernquist92}
{Hernquist} L.,  {Weinberg} M.~D.,  1992, \mn@doi [\apj] {10.1086/171975},
  \href {http://adsabs.harvard.edu/abs/1992ApJ...400...80H} {400, 80}

\bibitem[\protect\citeauthoryear{Holley-Bockelmann, Weinberg  \&
  Katz}{Holley-Bockelmann et~al.}{2005}]{holleyb05}
Holley-Bockelmann K.,  Weinberg M.,   Katz N.,  2005, MNRAS, 363, 991

\bibitem[\protect\citeauthoryear{{Kim} et~al.,}{{Kim} et~al.}{2014}]{kim14}
{Kim} T.,  et~al., 2014, \mn@doi [\apj] {10.1088/0004-637X/782/2/64}, \href
  {http://adsabs.harvard.edu/abs/2014ApJ...782...64K} {782, 64}

\bibitem[\protect\citeauthoryear{{Kormendy} \& {Kennicutt}}{{Kormendy} \&
  {Kennicutt}}{2004}]{kormendy04}
{Kormendy} J.,  {Kennicutt} Jr. R.~C.,  2004, \mn@doi [\araa]
  {10.1146/annurev.astro.42.053102.134024}, \href
  {http://adsabs.harvard.edu/abs/2004ARA%26A..42..603K} {42, 603}

\bibitem[\protect\citeauthoryear{{Kravtsov}}{{Kravtsov}}{2013}]{kravtsov13}
{Kravtsov} A.~V.,  2013, \mn@doi [\apjl] {10.1088/2041-8205/764/2/L31}, \href
  {http://adsabs.harvard.edu/abs/2013ApJ...764L..31K} {764, L31}

\bibitem[\protect\citeauthoryear{{Kuijken} \& {Dubinski}}{{Kuijken} \&
  {Dubinski}}{1994}]{kuijken94}
{Kuijken} K.,  {Dubinski} J.,  1994, \mnras, \href
  {http://adsabs.harvard.edu/abs/1994MNRAS.269...13K} {269, 13}

\bibitem[\protect\citeauthoryear{{Laurikainen}, {Salo}, {Athanassoula}, {Bosma}
   \& {Herrera-Endoqui}}{{Laurikainen} et~al.}{2014}]{laurikainen14}
{Laurikainen} E.,  {Salo} H.,  {Athanassoula} E.,  {Bosma} A.,
  {Herrera-Endoqui} M.,  2014, \mn@doi [\mnras] {10.1093/mnrasl/slu118}, \href
  {http://adsabs.harvard.edu/abs/2014MNRAS.444L..80L} {444, L80}

\bibitem[\protect\citeauthoryear{{Ling}}{{Ling}}{2010}]{ling10}
{Ling} F.-S.,  2010, \mn@doi [\prd] {10.1103/PhysRevD.82.023534}, \href
  {http://adsabs.harvard.edu/abs/2010PhRvD..82b3534L} {82, 023534}

\bibitem[\protect\citeauthoryear{Lloyd}{Lloyd}{1982}]{lloyd82}
Lloyd S.,  1982, IEEE Transactions on Information Theory, 28, 2

\bibitem[\protect\citeauthoryear{{Lynden-Bell} \& {Kalnajs}}{{Lynden-Bell} \&
  {Kalnajs}}{1972}]{lynden72}
{Lynden-Bell} D.,  {Kalnajs} A.~J.,  1972, \mnras, \href
  {http://adsabs.harvard.edu/abs/1972MNRAS.157....1L} {157, 1}

\bibitem[\protect\citeauthoryear{{Martinez-Valpuesta}, {Shlosman}  \&
  {Heller}}{{Martinez-Valpuesta} et~al.}{2006}]{martinez06}
{Martinez-Valpuesta} I.,  {Shlosman} I.,   {Heller} C.,  2006, \mn@doi [\apj]
  {10.1086/498338}, \href {http://adsabs.harvard.edu/abs/2006ApJ...637..214M}
  {637, 214}

\bibitem[\protect\citeauthoryear{{Martinsson}, {Verheijen}, {Westfall},
  {Bershady}, {Andersen}  \& {Swaters}}{{Martinsson}
  et~al.}{2013}]{martinsson13}
{Martinsson} T.~P.~K.,  {Verheijen} M.~A.~W.,  {Westfall} K.~B.,  {Bershady}
  M.~A.,  {Andersen} D.~R.,   {Swaters} R.~A.,  2013, \mn@doi [\aap]
  {10.1051/0004-6361/201321390}, \href
  {http://adsabs.harvard.edu/abs/2013A%26A...557A.131M} {557, A131}

\bibitem[\protect\citeauthoryear{{Mashchenko}, {Couchman}  \&
  {Wadsley}}{{Mashchenko} et~al.}{2006}]{mashchenko06}
{Mashchenko} S.,  {Couchman} H.~M.~P.,   {Wadsley} J.,  2006, \mn@doi [\nat]
  {10.1038/nature04944}, \href
  {http://adsabs.harvard.edu/abs/2006Natur.442..539M} {442, 539}

\bibitem[\protect\citeauthoryear{{Masters} et~al.,}{{Masters}
  et~al.}{2012}]{masters12}
{Masters} K.~L.,  et~al., 2012, \mn@doi [\mnras]
  {10.1111/j.1365-2966.2012.21377.x}, \href
  {http://adsabs.harvard.edu/abs/2012MNRAS.424.2180M} {424, 2180}

\bibitem[\protect\citeauthoryear{{Moster}, {Naab}  \& {White}}{{Moster}
  et~al.}{2013}]{moster13}
{Moster} B.~P.,  {Naab} T.,   {White} S.~D.~M.,  2013, \mn@doi [\mnras]
  {10.1093/mnras/sts261}, \href
  {http://adsabs.harvard.edu/abs/2013MNRAS.428.3121M} {428, 3121}

\bibitem[\protect\citeauthoryear{{Mu{\~n}oz-Mateos} et~al.,}{{Mu{\~n}oz-Mateos}
  et~al.}{2013}]{munoz13}
{Mu{\~n}oz-Mateos} J.~C.,  et~al., 2013, \mn@doi [\apj]
  {10.1088/0004-637X/771/1/59}, \href
  {http://adsabs.harvard.edu/abs/2013ApJ...771...59M} {771, 59}

\bibitem[\protect\citeauthoryear{{Navarro}, {Frenk}  \& {White}}{{Navarro}
  et~al.}{1997}]{navarro97}
{Navarro} J.~F.,  {Frenk} C.~S.,   {White} S.~D.~M.,  1997, \apj, \href
  {http://adsabs.harvard.edu/abs/1997ApJ...490..493N} {490, 493}

\bibitem[\protect\citeauthoryear{{Petersen}, {Katz}  \& {Weinberg}}{{Petersen}
  et~al.}{2016}]{petersen15a}
{Petersen} M.,  {Katz} N.,   {Weinberg} M.,  2016, \prd

\bibitem[\protect\citeauthoryear{{Piffl} et~al.,}{{Piffl}
  et~al.}{2014}]{piffl14a}
{Piffl} T.,  et~al., 2014, \mn@doi [\mnras] {10.1093/mnras/stu1948}, \href
  {http://adsabs.harvard.edu/abs/2014MNRAS.445.3133P} {445, 3133}

\bibitem[\protect\citeauthoryear{{Pillepich}, {Kuhlen}, {Guedes}  \&
  {Madau}}{{Pillepich} et~al.}{2014}]{pillepich14}
{Pillepich} A.,  {Kuhlen} M.,  {Guedes} J.,   {Madau} P.,  2014, \mn@doi [\apj]
  {10.1088/0004-637X/784/2/161}, \href
  {http://adsabs.harvard.edu/abs/2014ApJ...784..161P} {784, 161}

\bibitem[\protect\citeauthoryear{Polyachenko \& Polyachenko}{Polyachenko \&
  Polyachenko}{1996}]{polyachenko96}
Polyachenko V.,  Polyachenko E.,  1996, AstL, 22, 302

\bibitem[\protect\citeauthoryear{{Purcell}, {Bullock}  \&
  {Kaplinghat}}{{Purcell} et~al.}{2009}]{purcell09}
{Purcell} C.~W.,  {Bullock} J.~S.,   {Kaplinghat} M.,  2009, \mn@doi [\apj]
  {10.1088/0004-637X/703/2/2275}, \href
  {http://adsabs.harvard.edu/abs/2009ApJ...703.2275P} {703, 2275}

\bibitem[\protect\citeauthoryear{{Read}, {Lake}, {Agertz}  \&
  {Debattista}}{{Read} et~al.}{2008}]{read08}
{Read} J.~I.,  {Lake} G.,  {Agertz} O.,   {Debattista} V.~P.,  2008, \mn@doi
  [\mnras] {10.1111/j.1365-2966.2008.13643.x}, \href
  {http://adsabs.harvard.edu/abs/2008MNRAS.389.1041R} {389, 1041}

\bibitem[\protect\citeauthoryear{{Read}, {Mayer}, {Brooks}, {Governato}  \&
  {Lake}}{{Read} et~al.}{2009}]{read09}
{Read} J.~I.,  {Mayer} L.,  {Brooks} A.~M.,  {Governato} F.,   {Lake} G.,
  2009, \mn@doi [\mnras] {10.1111/j.1365-2966.2009.14757.x}, \href
  {http://adsabs.harvard.edu/abs/2009MNRAS.397...44R} {397, 44}

\bibitem[\protect\citeauthoryear{{Rojas-Arriagada} et~al.,}{{Rojas-Arriagada}
  et~al.}{2014}]{rojas14}
{Rojas-Arriagada} A.,  et~al., 2014, \mn@doi [\aap]
  {10.1051/0004-6361/201424121}, \href
  {http://adsabs.harvard.edu/abs/2014A%26A...569A.103R} {569, A103}

\bibitem[\protect\citeauthoryear{{Rossi}, {Ortolani}, {Barbuy}, {Bica}  \&
  {Bonfanti}}{{Rossi} et~al.}{2015}]{rossi15}
{Rossi} L.~J.,  {Ortolani} S.,  {Barbuy} B.,  {Bica} E.,   {Bonfanti} A.,
  2015, \mn@doi [\mnras] {10.1093/mnras/stv748}, \href
  {http://adsabs.harvard.edu/abs/2015MNRAS.450.3270R} {450, 3270}

\bibitem[\protect\citeauthoryear{{Saha} \& {Naab}}{{Saha} \&
  {Naab}}{2013}]{saha13}
{Saha} K.,  {Naab} T.,  2013, \mn@doi [\mnras] {10.1093/mnras/stt1088}, \href
  {http://adsabs.harvard.edu/abs/2013MNRAS.434.1287S} {434, 1287}

\bibitem[\protect\citeauthoryear{{Saha}, {Tseng}  \& {Taam}}{{Saha}
  et~al.}{2010}]{saha10}
{Saha} K.,  {Tseng} Y.-H.,   {Taam} R.~E.,  2010, \mn@doi [\apj]
  {10.1088/0004-637X/721/2/1878}, \href
  {http://adsabs.harvard.edu/abs/2010ApJ...721.1878S} {721, 1878}

\bibitem[\protect\citeauthoryear{{Saha}, {Martinez-Valpuesta}  \&
  {Gerhard}}{{Saha} et~al.}{2012}]{saha12}
{Saha} K.,  {Martinez-Valpuesta} I.,   {Gerhard} O.,  2012, \mn@doi [\mnras]
  {10.1111/j.1365-2966.2011.20307.x}, \href
  {http://adsabs.harvard.edu/abs/2012MNRAS.421..333S} {421, 333}

\bibitem[\protect\citeauthoryear{Sellwood}{Sellwood}{2014}]{sellwood14rev}
Sellwood J.,  2014, RevModPhys, 86, 1

\bibitem[\protect\citeauthoryear{{Sellwood} \& {Debattista}}{{Sellwood} \&
  {Debattista}}{2006}]{sellwood06b}
{Sellwood} J.~A.,  {Debattista} V.~P.,  2006, \mn@doi [\apj] {10.1086/499482},
  \href {http://adsabs.harvard.edu/abs/2006ApJ...639..868S} {639, 868}

\bibitem[\protect\citeauthoryear{{Sellwood} \& {Debattista}}{{Sellwood} \&
  {Debattista}}{2009}]{sellwood09}
{Sellwood} J.~A.,  {Debattista} V.~P.,  2009, \mn@doi [\mnras]
  {10.1111/j.1365-2966.2009.15219.x}, \href
  {http://adsabs.harvard.edu/abs/2009MNRAS.398.1279S} {398, 1279}

\bibitem[\protect\citeauthoryear{{Sellwood} \& {Wilkinson}}{{Sellwood} \&
  {Wilkinson}}{1993}]{sellwood93}
{Sellwood} J.~A.,  {Wilkinson} A.,  1993, \mn@doi [Reports on Progress in
  Physics] {10.1088/0034-4885/56/2/001}, \href
  {http://adsabs.harvard.edu/abs/1993RPPh...56..173S} {56, 173}

\bibitem[\protect\citeauthoryear{{Sheth} et~al.,}{{Sheth}
  et~al.}{2008}]{sheth08}
{Sheth} K.,  et~al., 2008, \mn@doi [\apj] {10.1086/524980}, \href
  {http://adsabs.harvard.edu/abs/2008ApJ...675.1141S} {675, 1141}

\bibitem[\protect\citeauthoryear{{Soto}, {Zeballos}, {Kuijken}, {Rich},
  {Kunder}  \& {Astraatmadja}}{{Soto} et~al.}{2014}]{soto14}
{Soto} M.,  {Zeballos} H.,  {Kuijken} K.,  {Rich} R.~M.,  {Kunder} A.,
  {Astraatmadja} T.,  2014, \mn@doi [\aap] {10.1051/0004-6361/201117339}, \href
  {http://adsabs.harvard.edu/abs/2014A%26A...562A..41S} {562, A41}

\bibitem[\protect\citeauthoryear{{Steinmetz} et~al.,}{{Steinmetz}
  et~al.}{2006}]{steinmetz06}
{Steinmetz} M.,  et~al., 2006, \mn@doi [\aj] {10.1086/506564}, \href
  {http://adsabs.harvard.edu/abs/2006AJ....132.1645S} {132, 1645}

\bibitem[\protect\citeauthoryear{{Tiret} \& {Combes}}{{Tiret} \&
  {Combes}}{2007}]{tiret07}
{Tiret} O.,  {Combes} F.,  2007, \mn@doi [\aap] {10.1051/0004-6361:20066446},
  \href {http://adsabs.harvard.edu/abs/2007A%26A...464..517T} {464, 517}

\bibitem[\protect\citeauthoryear{{Tremaine} \& {Ostriker}}{{Tremaine} \&
  {Ostriker}}{1999}]{tremaine99}
{Tremaine} S.,  {Ostriker} J.~P.,  1999, \mn@doi [\mnras]
  {10.1046/j.1365-8711.1999.02558.x}, \href
  {http://adsabs.harvard.edu/abs/1999MNRAS.306..662T} {306, 662}

\bibitem[\protect\citeauthoryear{Tremaine \& Weinberg}{Tremaine \&
  Weinberg}{1984}]{tremaine84}
Tremaine S.,  Weinberg M.,  1984, MNRAS, 209, 729

\bibitem[\protect\citeauthoryear{{Villa-Vargas}, {Shlosman}  \&
  {Heller}}{{Villa-Vargas} et~al.}{2009}]{villavargas09}
{Villa-Vargas} J.,  {Shlosman} I.,   {Heller} C.,  2009, \mn@doi [\apj]
  {10.1088/0004-637X/707/1/218}, \href
  {http://adsabs.harvard.edu/abs/2009ApJ...707..218V} {707, 218}

\bibitem[\protect\citeauthoryear{{Villa-Vargas}, {Shlosman}  \&
  {Heller}}{{Villa-Vargas} et~al.}{2010}]{villavargas10}
{Villa-Vargas} J.,  {Shlosman} I.,   {Heller} C.,  2010, \mn@doi [\apj]
  {10.1088/0004-637X/719/2/1470}, \href
  {http://adsabs.harvard.edu/abs/2010ApJ...719.1470V} {719, 1470}

\bibitem[\protect\citeauthoryear{{Vogelsberger} et~al.,}{{Vogelsberger}
  et~al.}{2014}]{vogelsberger14}
{Vogelsberger} M.,  et~al., 2014, \mn@doi [\mnras] {10.1093/mnras/stu1536},
  \href {http://adsabs.harvard.edu/abs/2014MNRAS.444.1518V} {444, 1518}

\bibitem[\protect\citeauthoryear{{Wegg}, {Gerhard}  \& {Portail}}{{Wegg}
  et~al.}{2015}]{wegg15}
{Wegg} C.,  {Gerhard} O.,   {Portail} M.,  2015, \mn@doi [\mnras]
  {10.1093/mnras/stv745}, \href
  {http://adsabs.harvard.edu/abs/2015MNRAS.450.4050W} {450, 4050}

\bibitem[\protect\citeauthoryear{{Weinberg}}{{Weinberg}}{1985}]{weinberg85}
{Weinberg} M.~D.,  1985, \mnras, \href
  {http://adsabs.harvard.edu/abs/1985MNRAS.213..451W} {213, 451}

\bibitem[\protect\citeauthoryear{{Weinberg}}{{Weinberg}}{1991}]{weinberg91}
{Weinberg} M.~D.,  1991, \mn@doi [\apj] {10.1086/169671}, \href
  {http://adsabs.harvard.edu/abs/1991ApJ...368...66W} {368, 66}

\bibitem[\protect\citeauthoryear{{Weinberg}}{{Weinberg}}{1994}]{weinberg94}
{Weinberg} M.~D.,  1994, \mn@doi [\apj] {10.1086/173665}, \href
  {http://adsabs.harvard.edu/abs/1994ApJ...421..481W} {421, 481}

\bibitem[\protect\citeauthoryear{{Weinberg}}{{Weinberg}}{1998}]{weinberg98}
{Weinberg} M.~D.,  1998, \mn@doi [\mnras] {10.1046/j.1365-8711.1998.01456.x},
  \href {http://adsabs.harvard.edu/abs/1998MNRAS.297..101W} {297, 101}

\bibitem[\protect\citeauthoryear{Weinberg}{Weinberg}{1999}]{weinberg99}
Weinberg M.,  1999, AJ, 117, 629

\bibitem[\protect\citeauthoryear{Weinberg \& Katz}{Weinberg \&
  Katz}{2002}]{weinberg02}
Weinberg M.,  Katz N.,  2002, ApJ, 580, 627

\bibitem[\protect\citeauthoryear{{Weinberg} \& {Katz}}{{Weinberg} \&
  {Katz}}{2007a}]{weinberg07a}
{Weinberg} M.~D.,  {Katz} N.,  2007a, \mn@doi [\mnras]
  {10.1111/j.1365-2966.2006.11306.x}, \href
  {http://adsabs.harvard.edu/abs/2007MNRAS.375..425W} {375, 425}

\bibitem[\protect\citeauthoryear{{Weinberg} \& {Katz}}{{Weinberg} \&
  {Katz}}{2007b}]{weinberg07b}
{Weinberg} M.~D.,  {Katz} N.,  2007b, \mn@doi [\mnras]
  {10.1111/j.1365-2966.2006.11307.x}, \href
  {http://adsabs.harvard.edu/abs/2007MNRAS.375..460W} {375, 460}

\bibitem[\protect\citeauthoryear{{Williams}, {Bureau}  \&
  {Kuntschner}}{{Williams} et~al.}{2012}]{williams12}
{Williams} M.~J.,  {Bureau} M.,   {Kuntschner} H.,  2012, \mn@doi [\mnras]
  {10.1111/j.1745-3933.2012.01353.x}, \href
  {http://adsabs.harvard.edu/abs/2012MNRAS.427L..99W} {427, L99}

\bibitem[\protect\citeauthoryear{{Zaritsky} et~al.,}{{Zaritsky}
  et~al.}{2013}]{zaritsky13}
{Zaritsky} D.,  et~al., 2013, \mn@doi [\apj] {10.1088/0004-637X/772/2/135},
  \href {http://adsabs.harvard.edu/abs/2013ApJ...772..135Z} {772, 135}

\makeatother
\end{thebibliography}

\end{document}